\documentclass[10pt,twocolumn,twoside,openbib]{article}
\usepackage{graphicx}
\usepackage{float}
\usepackage[footnotesize]{caption}
\newcommand{\icar}[1]{}
\newcommand{\prep}[1]{#1}
\prep{\setlength{\textwidth}{ 18cm}}
\prep{\setlength{\textheight}{  22.cm}}

\icar{\setlength{\textwidth}{ 16cm}}
\icar{\setlength{\textheight}{ 20.cm}}
\def \eps {\varepsilon}

\def\etal{{\it et al.}}
\def\llabel#1{\label{#1}} 
 
\def\cS{{\cal S}}
\def\abs#1{\left\vert#1\right\vert}
\def\syr{''/\hbox{yr}}

\def\m@th{\mathsurround=0pt}
\def\EQM#1{\vcenter{\normalbaselines\m@th
    \ialign{${\displaystyle ##}$\hfil&&\ ${\displaystyle ##}$\hfil\crcr
    \mathstrut\crcr\noalign{\kern-\baselineskip}
    \noalign{\smallskip}
    #1\crcr\mathstrut\crcr\noalign{\kern-\baselineskip}}}}
\newcommand{\Frac}[2]{{{\displaystyle\strut#1}\over{\displaystyle\strut#2}}}
\newcommand{\be}{\begin{equation}}

\newcommand{\ee}{\end{equation}}
\prep{\def\figsize{8.cm}}
\prep{}
\prep{\def\figSIZE{13.cm}}
\icar{\def\figsize{11.cm}}
\icar{}
\icar{\def\figSIZE{16.cm}}

\newcommand\figNa{
\begin{figure*}[t] 
 \includegraphics[width=\figSIZE]{figures_pdf/res_ecc_tot} 
  \caption{Normalized density functions for the eccentricities of the  planets. The statistic is established 
  with 1001 solutions with very close  initial conditions. The evolution is followed over 19 intervals of 250 Myr,
  represented by the different curves. The variation of these curves thus reflects the 
  chaotic diffusion of the solutions. For the outer planets, all 19 curves are practically 
  identical.
  } 
  \llabel{FigNa}
\end{figure*}
}

\newcommand\figNb{
\begin{figure*}[t] 
 \includegraphics[width=\figSIZE]{figures_pdf/res_inc_tot} 
  \caption{Normalized density functions for the inclinations of the  planets. The statistic is established 
  with 1001 solutions with very close  initial conditions. The evolution is followed over 19 intervals of 250 Myr,
  represented by the different curves. The variation of these curves thus reflects the 
  chaotic diffusion of the solutions. For the outer planets, all 19 curves are practically 
  identical.
  } 
  \llabel{FigNb}
\end{figure*}
}

\newcommand\figNc{
\begin{figure*}[t] 
 \includegraphics[width=12.5cm]{figures_pdf/maxecc_incC} 
  \caption{Probability (in percents) to reach a given value of the eccentricity (left) or inclination (right) 
  over a given time interval. The statistic is established 
  with 1001 solution with very close  initial conditions. The maximum values are computed over 
  50, 100, 250, 500, 1000, 2000, 3000, 4000, and 5000 Myr for the inner planets. For the outer planets, the diffusion is so small 
  that these computations are easily summarized in a Table. Inclinations are computed with respect 
  to the ecliptic J2000.
  } 
  \llabel{FigNc}
\end{figure*}
}

\newcommand\figNd{
\begin{figure}[h] 
\begin{center}
 \includegraphics[width=\figsize]{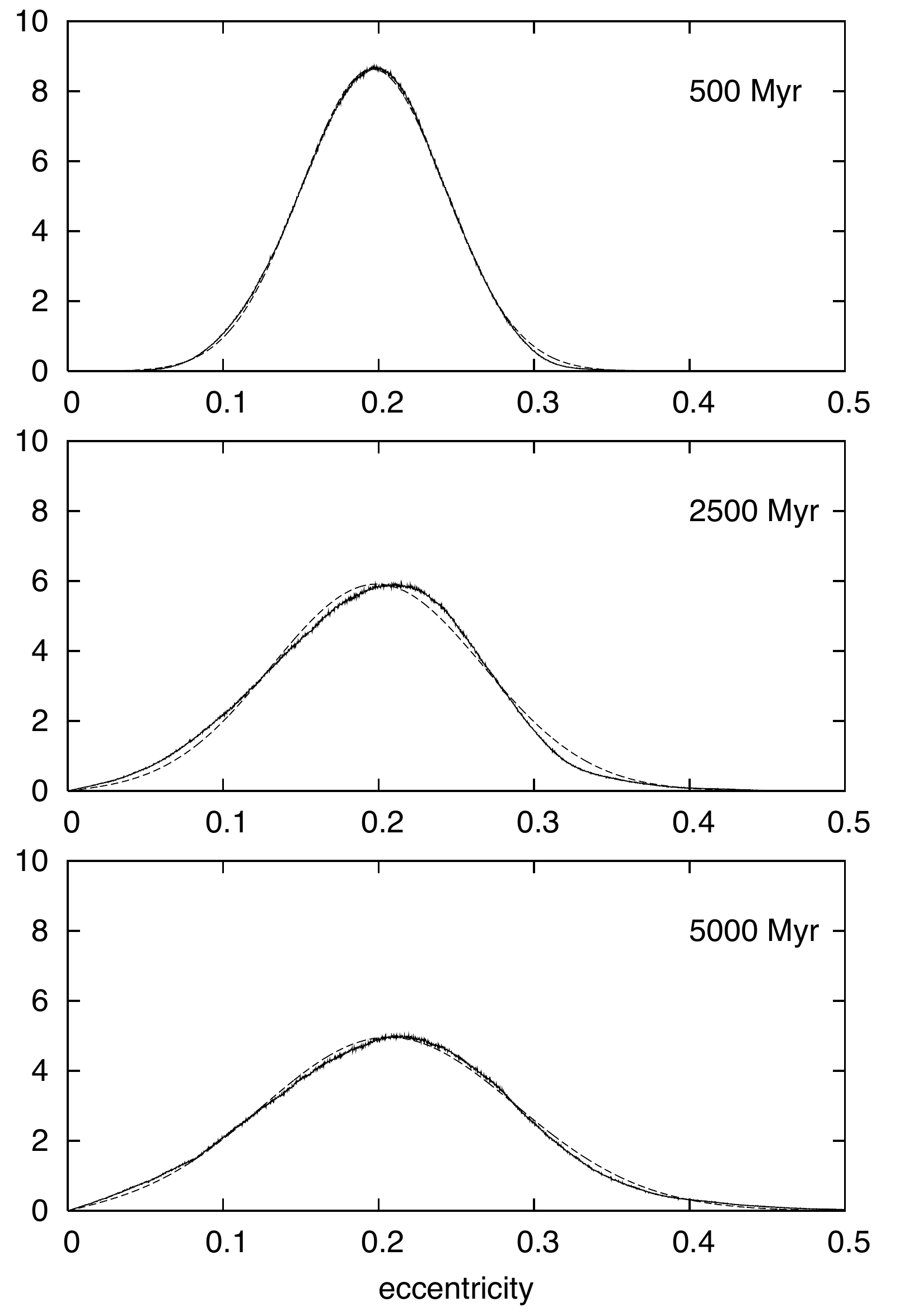} 
  \caption{PDF of the eccentricity of  Mercury (full line) for three different 
   dates from J2000 (500 Myr, 2500 Myr, 5000 Myr). For each case, a two parameters 
   Rice PDF (dashed line) has been fitted to the eccentricity PDF.}
  \llabel{FigNd}
\end{center}
\end{figure}
} 

\newcommand\figNe{
\begin{figure}[h] 
\begin{center}
 \includegraphics[width=\figsize]{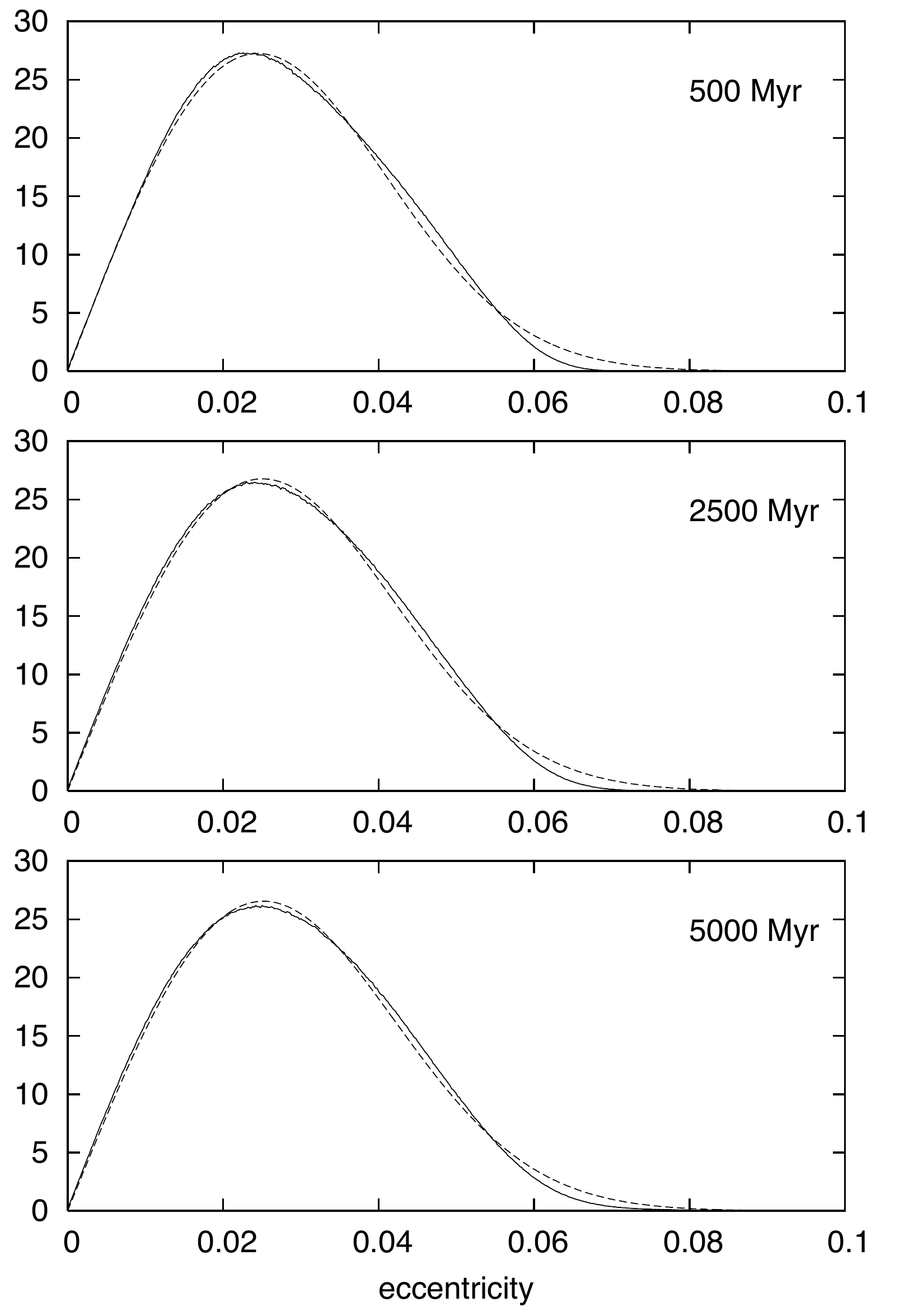} 
  \caption{PDF of the eccentricity of  the Earth (full line) for three different 
   dates from J2000 (500 Myr, 2500 Myr, 5000 Myr). For each case, a two parameters 
   Rice PDF (dashed line) has been fitted to the eccentricity PDF.}
  \llabel{FigNe}
\end{center}
\end{figure}
} 

\newcommand\figNeb{
\begin{figure}[h] 
\begin{center}
 \includegraphics[width=\figsize]{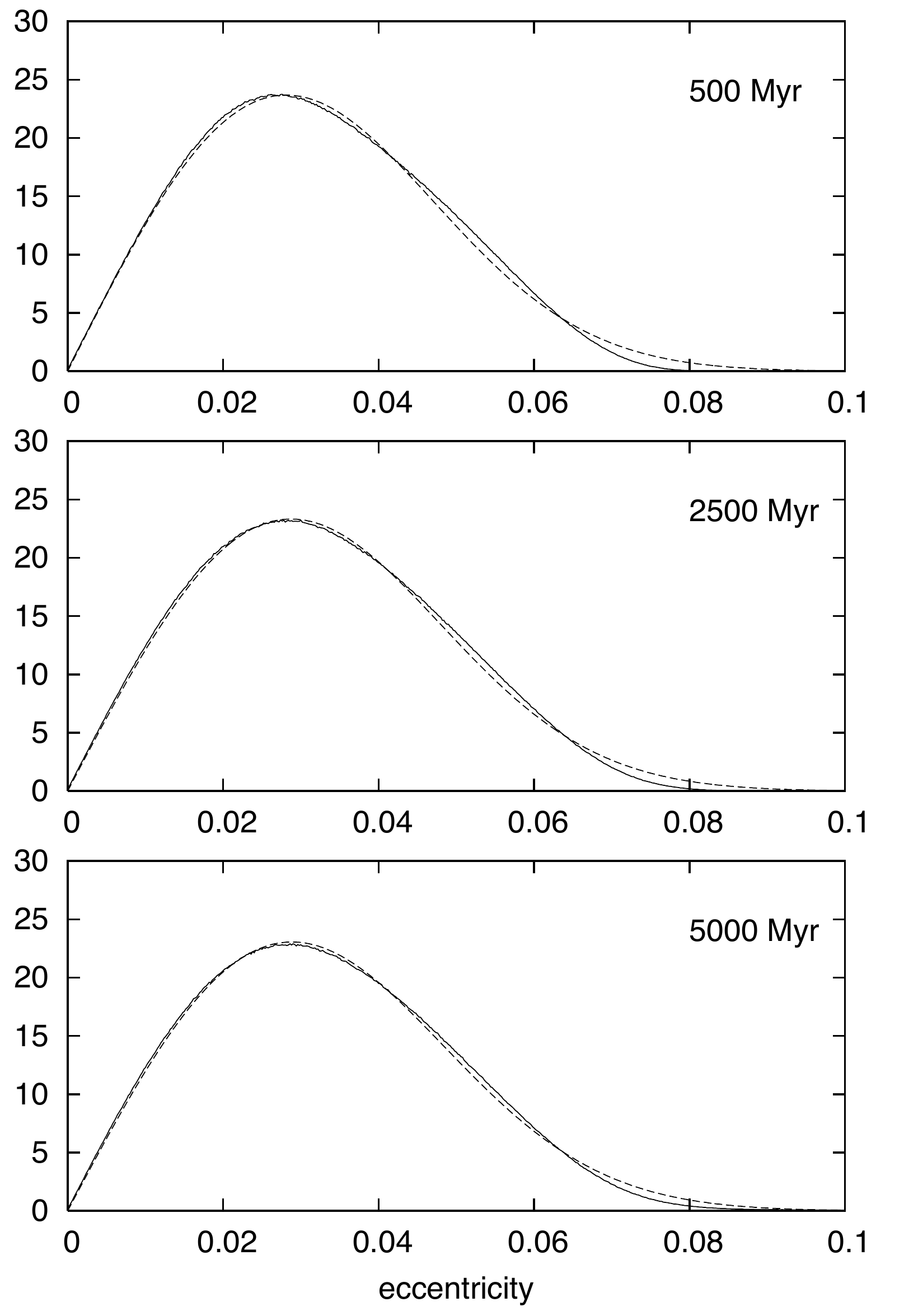} 
  \caption{PDF of the eccentricity of  Venus (full line) for three different 
   dates from J2000 (500 Myr, 2500 Myr, 5000 Myr). For each case, a two parameters 
   Rice PDF (dashed line) has been fitted to the eccentricity PDF.}
  \llabel{FigNeb}
\end{center}
\end{figure}
} 

\newcommand\figNf{
\begin{figure}[h] 
\begin{center}
 \includegraphics[width=\figsize]{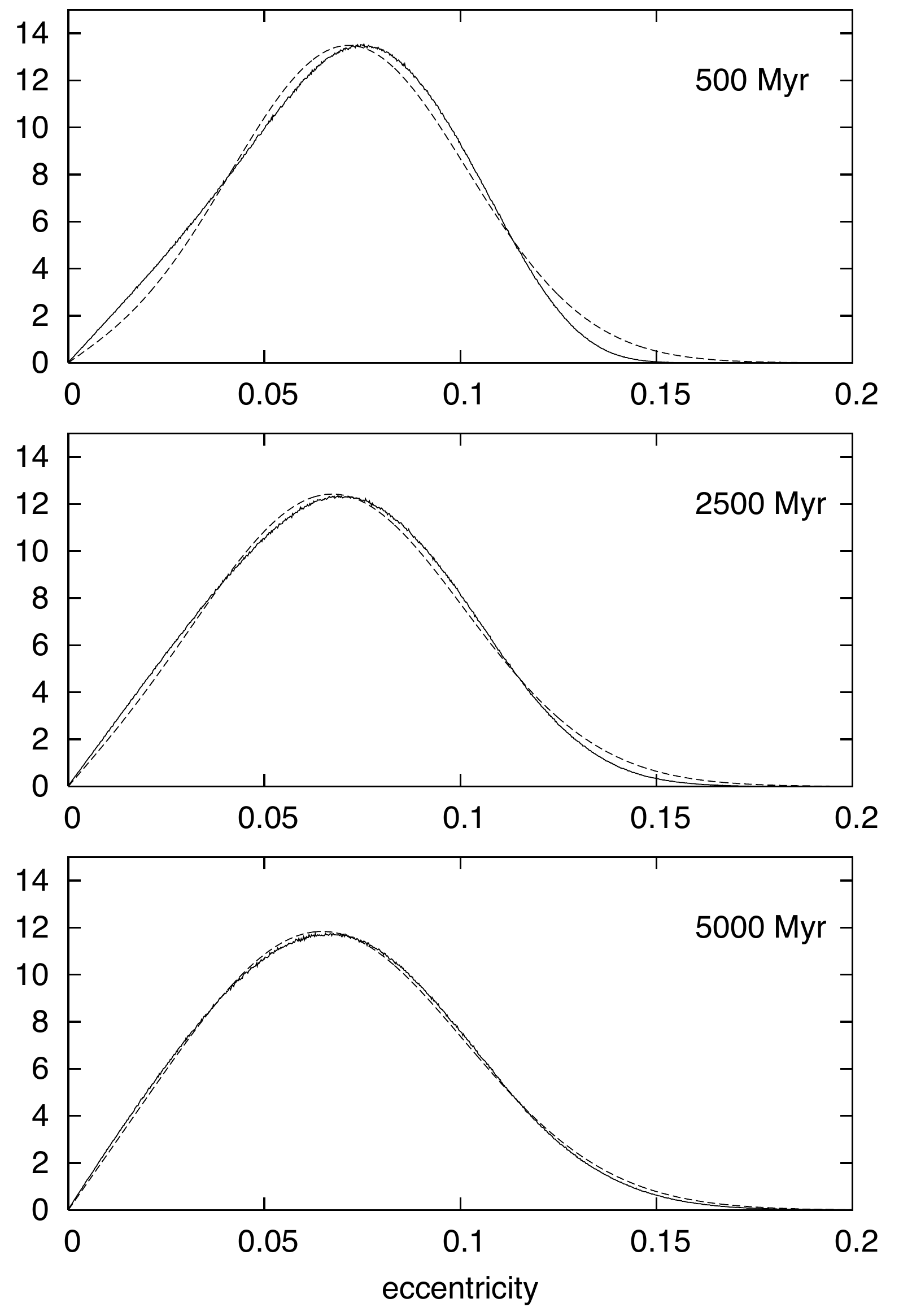} 
  \caption{PDF of the eccentricity of  Mars (full line) for three different 
   dates from J2000 (500 Myr, 2500 Myr, 5000 Myr). For each case, a two parameters 
   Rice PDF (dashed line) has been fitted to the eccentricity PDF.}
  \llabel{FigNf}
\end{center}
\end{figure}
}

\newcommand\figNg{
\begin{figure}[h] 
\begin{center}
 \includegraphics[width=\figsize]{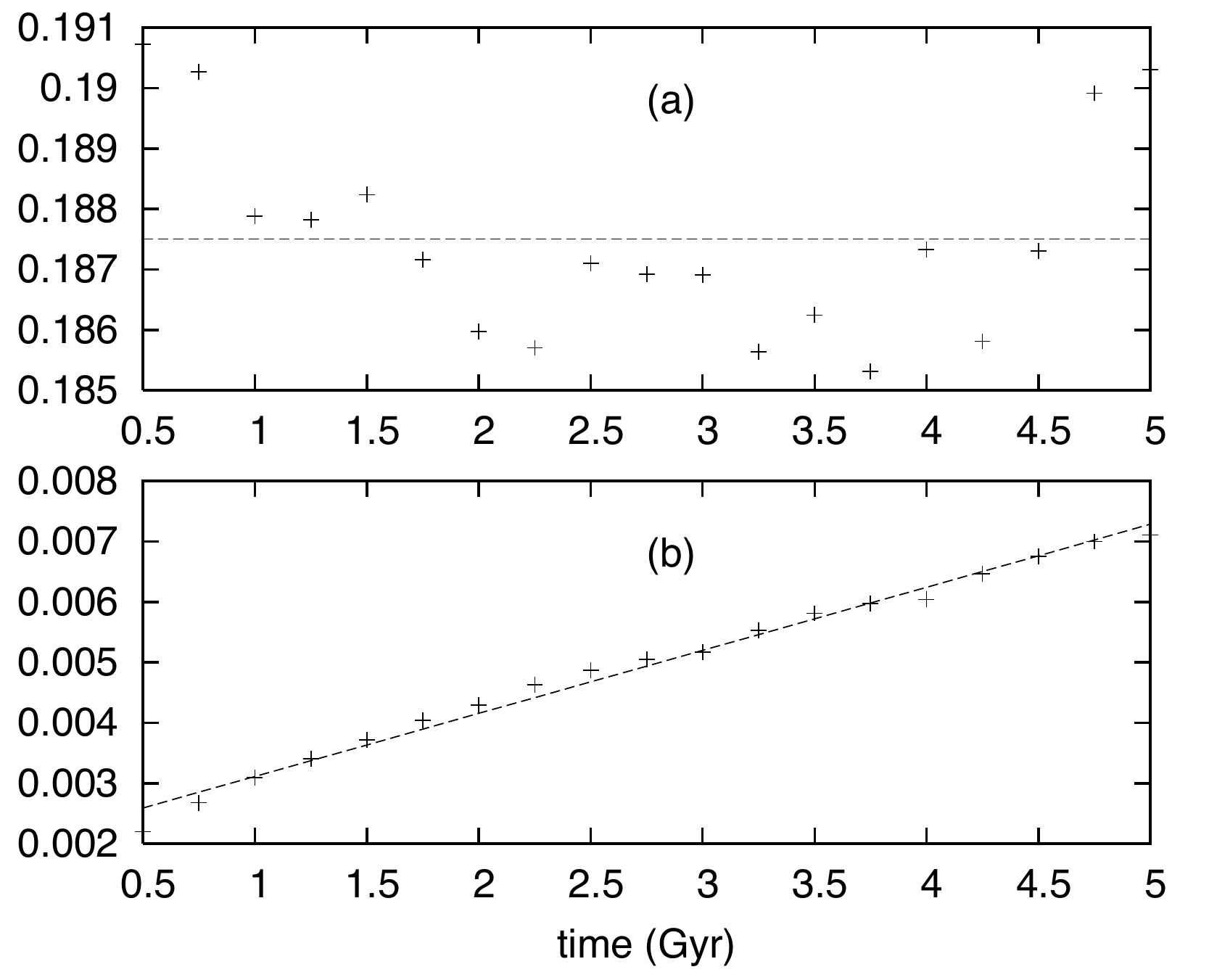} 
  \caption{ Evolution with time of the parameters $m$ (top) and $b$ (bottom) 
  of the PDF of the eccentricity of Mercury. $m$  is approximmated by  a constant value $m_0$  (dotted line), while $b$
  is fitted with a linear slope  $b= b_0 +  b_1 T$, where the time $T$ is in Gyr.
  The fitted values $m_0, b_0, b_1$ are given in Table \ref{TabNa}.
  }
  \llabel{FigNg}
\end{center}
\end{figure}
}

\newcommand\figNhb{
\begin{figure}[h] 
\begin{center}
 \includegraphics[width=\figsize]{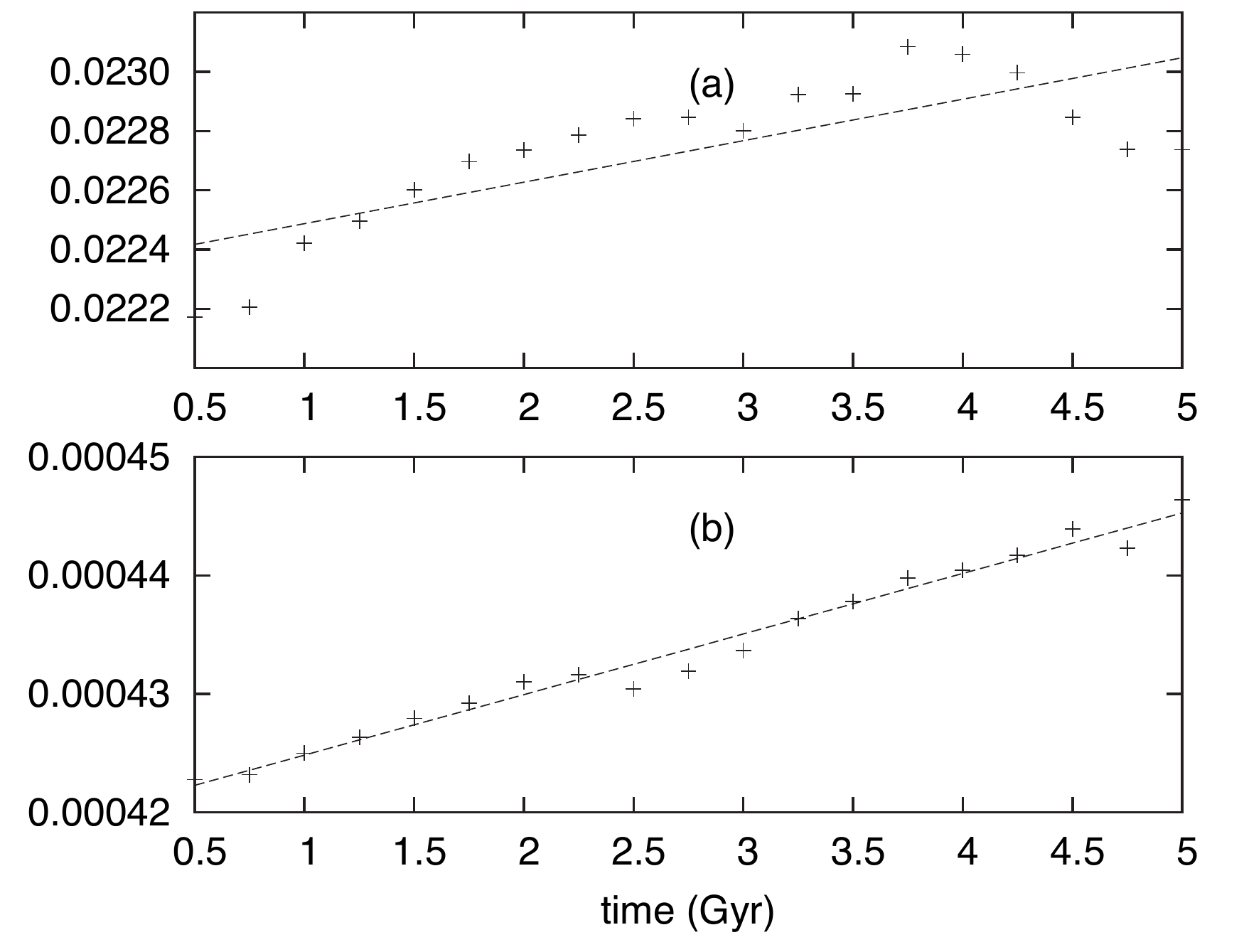} 
  \caption{ Evolution with time of the parameters $m$ (top) and $b$ (bottom) 
  of the PDF of the eccentricity of Venus. $m$ and $b$
  are fitted with a linear slope  $m=m_0+m_1 T$ and  $b= b_0 +  b_1 T$  respectively, where the time $T$ is in Gyr.
  The fitted values $m_0, m_1, b_0, b_1$ are given in Table \ref{TabNa}.
  }
  \llabel{FigNhb}
\end{center}
\end{figure}
}

\newcommand\figNh{
\begin{figure}[h] 
\begin{center}
 \includegraphics[width=\figsize]{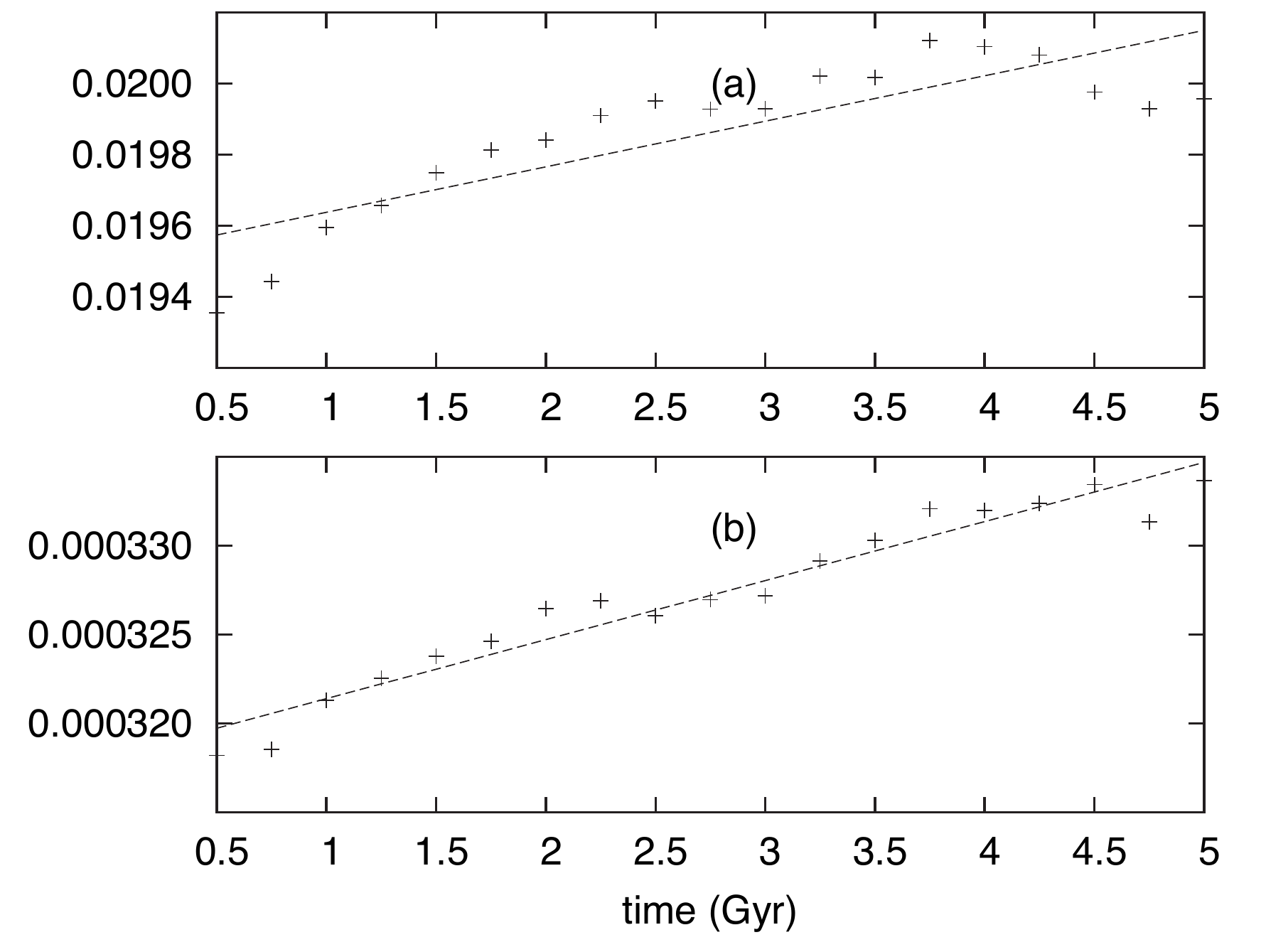} 
  \caption{ Evolution with time of the parameters $m$ (top) and $b$ (bottom) 
  of the PDF of the eccentricity of the Earth. $m$ and $b$
  are fitted with a linear slope  $m=m_0+m_1 T$ and   $b= b_0 +  b_1 T$  respectively, where the time $T$ is in Gyr.
  The fitted values $m_0, m_1, b_0, b_1$ are given in Table \ref{TabNa}.
  }
  \llabel{FigNh}
\end{center}
\end{figure}
}

\newcommand\figNi{
\begin{figure}[h] 
\begin{center}
 \includegraphics[width=\figsize]{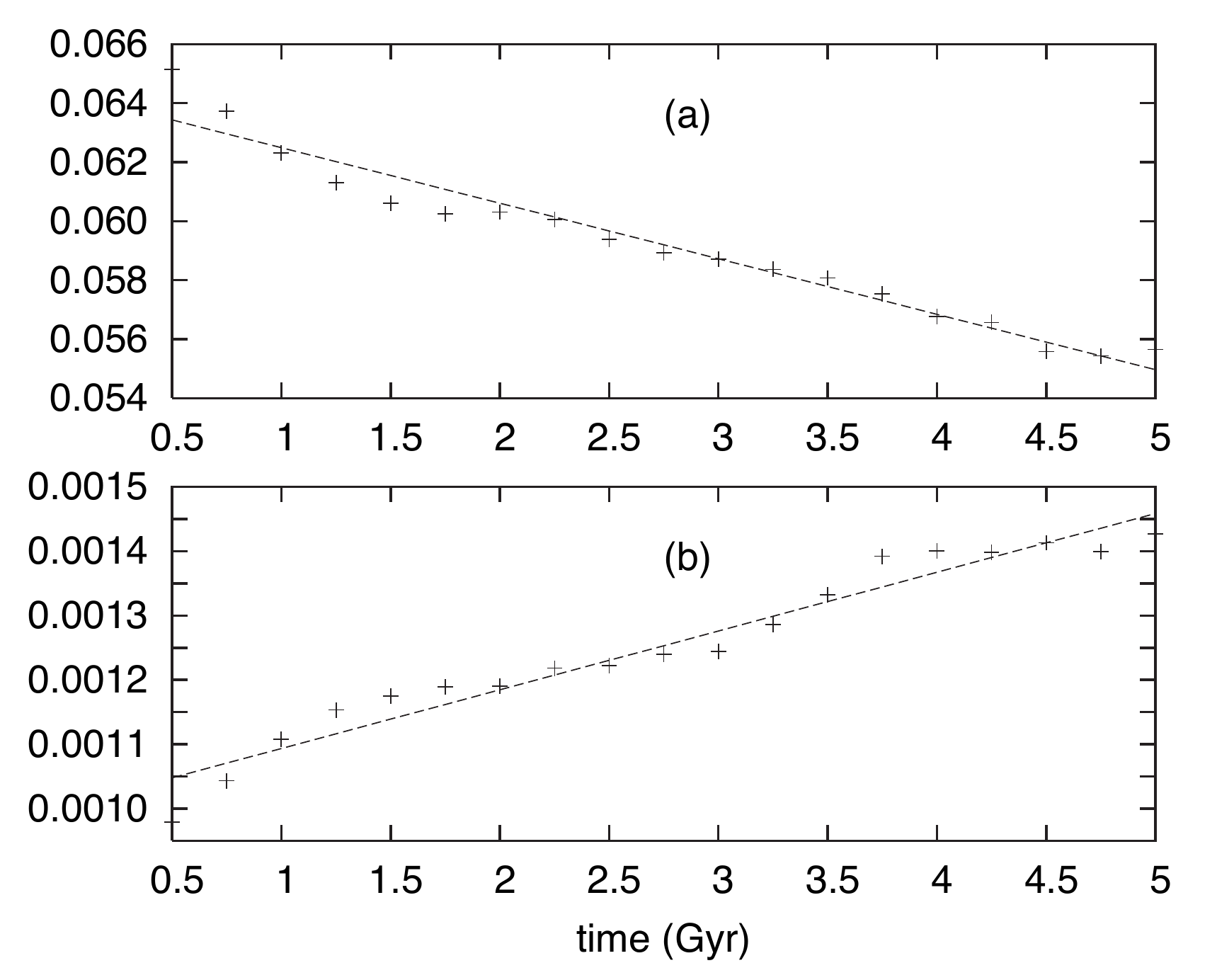} 
  \caption{ Evolution with time of the parameters $m$ (top) and $b$ (bottom) 
  of the PDF of the eccentricity of Mars. $m$  and $b$
  are fitted with a linear slope $m=m_0+m_1 T$ and  $b= b_0 +  b_1 T$  respectively, where the time $T$ is in Gyr.
  The fitted values $m_0, m_1, b_0, b_1$ are given in Table \ref{TabNa}.
  }
  \llabel{FigNi}
\end{center}
\end{figure}
}

\newcommand\figNj{
\begin{figure}[h] 
\begin{center}
 \includegraphics[width=\figsize]{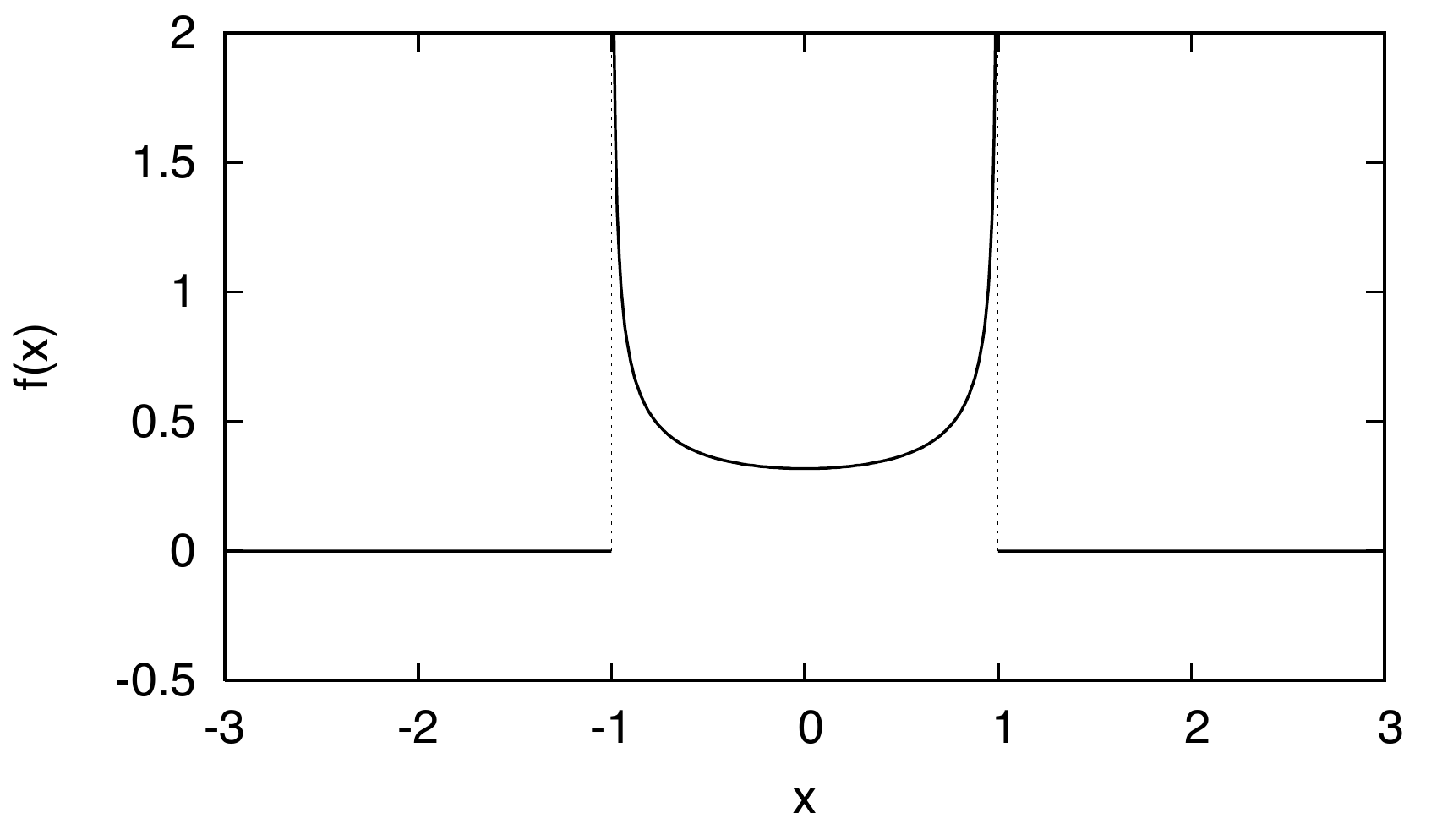} 
  \caption{Plot of the function $f(x) =1_{[-1,1]}(x)/(\pi\sqrt{1-x^2})$ 
  that is the PDF of $g(t)=\sin(t)$ (Eq.\ref{quasi5}).
  }
  \llabel{FigNj}
\end{center}
\end{figure}
}

\newcommand\figNk{
\begin{figure}[h] 
\begin{center}
 \includegraphics[width=8.5cm]{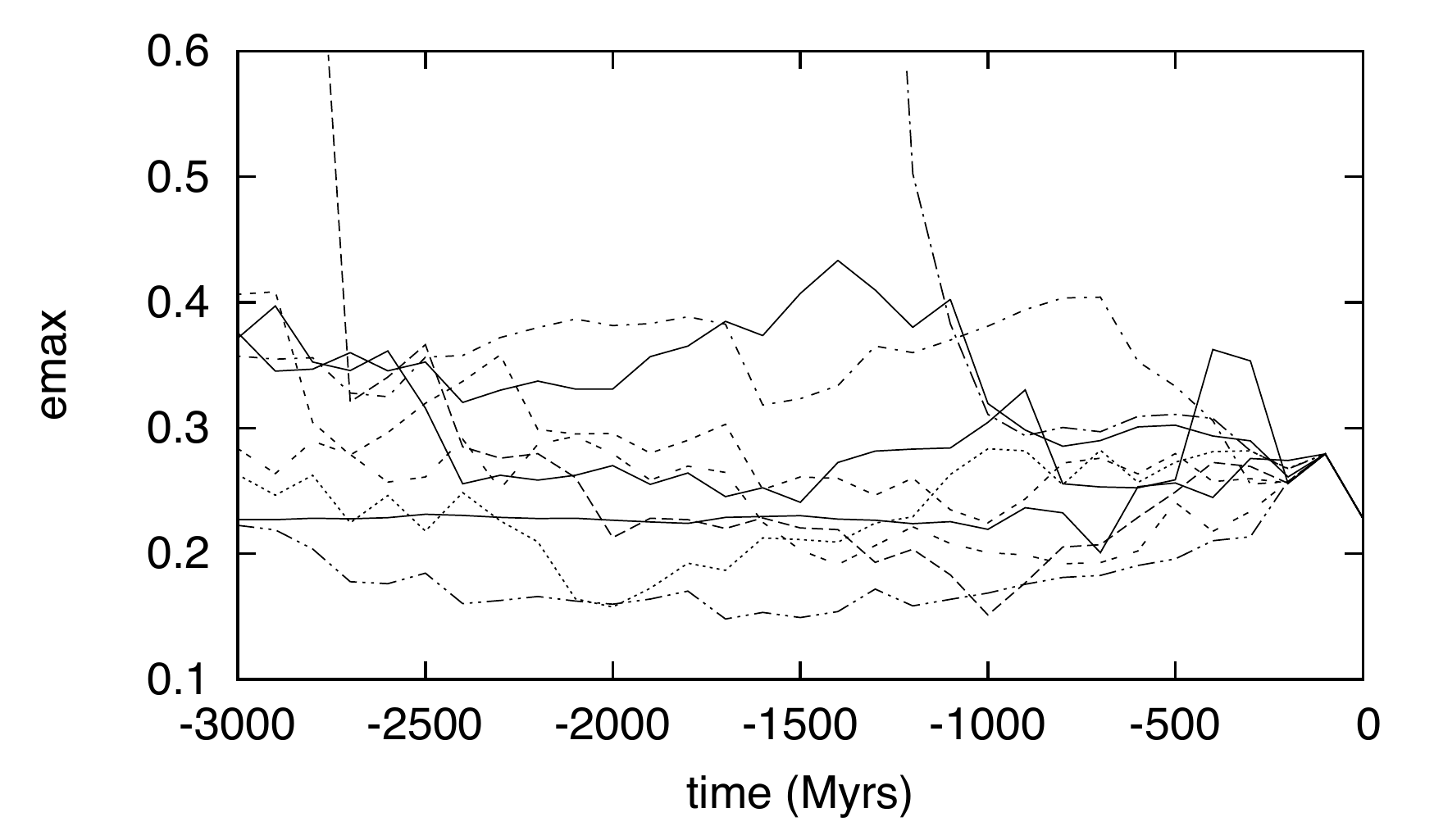} 
  \caption{Maximum value reached by the eccentricity of Mercury over 3Gyr. 
  The maximum values (emax) are computed over 500 Myr intervals for the 10 solutions 
  ${\cal S}_1,\dots {\cal S}_{10}$.
  }
  \llabel{FigNk}
\end{center}
\end{figure}
}
\newcommand\figNl{
\begin{figure}[h] 
\begin{center}
 \includegraphics[width=8.5cm]{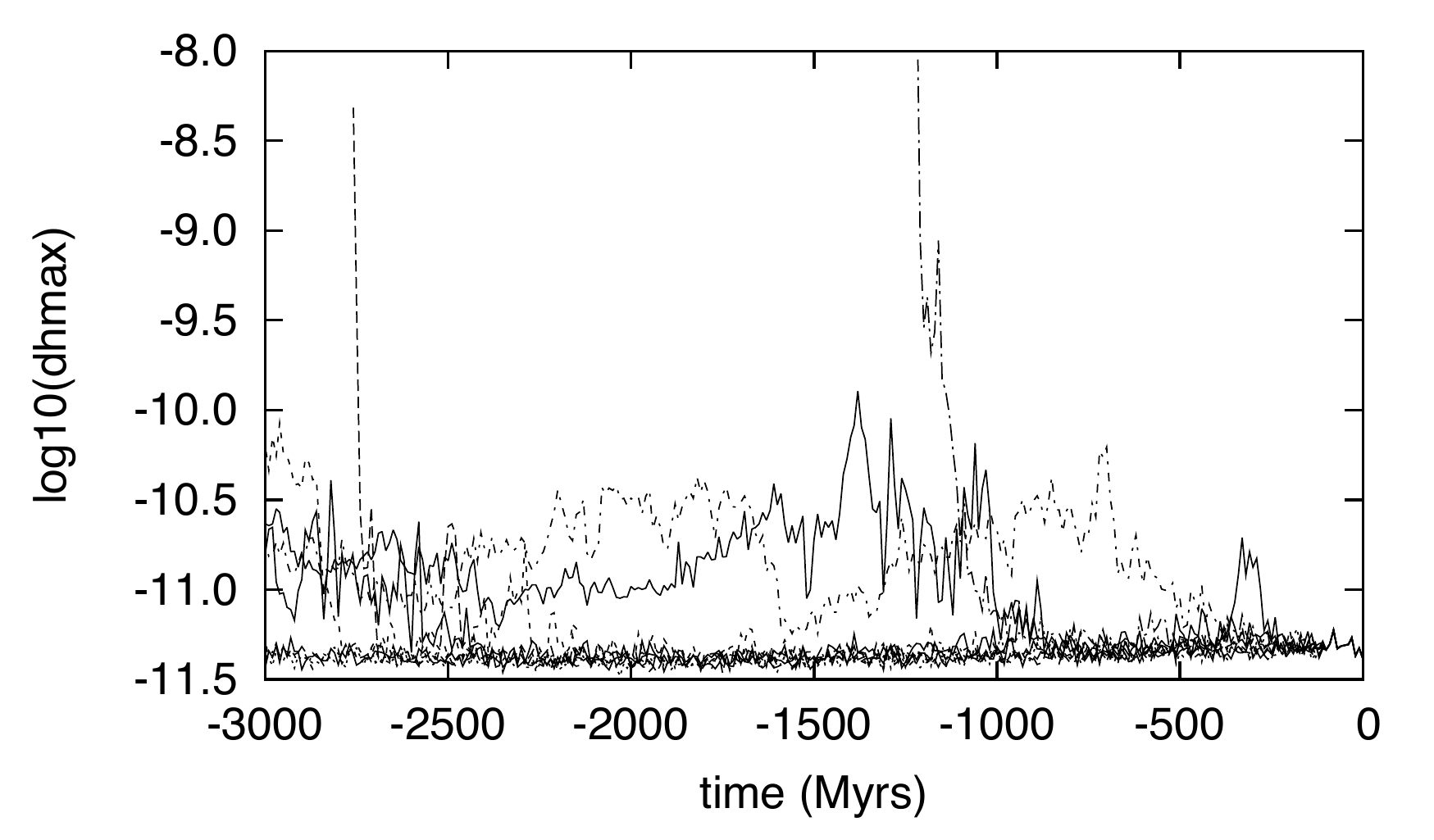} 
  \caption{Maximum value reached by the relative variation of the total energy ($\abs{(h(t)-h(0))/h(0)}$)
  of the Solar System for the 10 solutions 
  ${\cal S}_1,\dots {\cal S}_{10}$.  
  The maximum values  (dhmax)  are computed over 10 Myr intervals.
  }
  \llabel{FigNl}
\end{center}
\end{figure}
}
\newcommand\figNm{
\begin{figure}[h] 
\begin{center}
 \includegraphics[width=8.5cm]{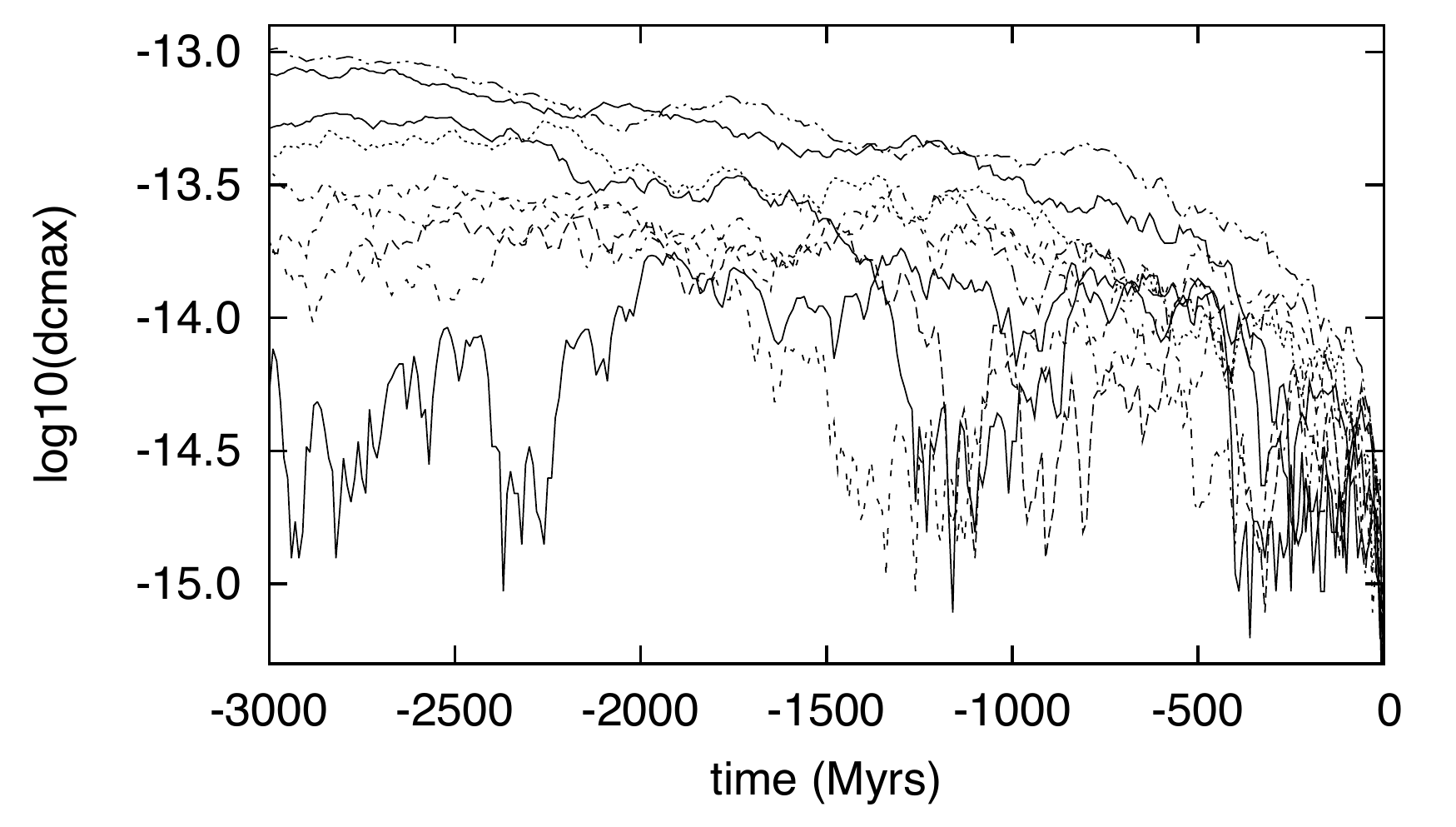} 
  \caption{Maximum value reached by the relative variation of the total angular momentum ($\abs{(c(t)-c(0))/c(0)}$)
  of the Solar System for the 10 solutions 
  ${\cal S}_1,\dots {\cal S}_{10}$.  
  The maximum values (dcmax) are computed over 10 Myr intervals.
  }
  \llabel{FigNm}
\end{center}
\end{figure}
}

\newcommand\figNn{
\begin{figure}[t] 
\begin{center}
 \includegraphics[width=8.5cm]{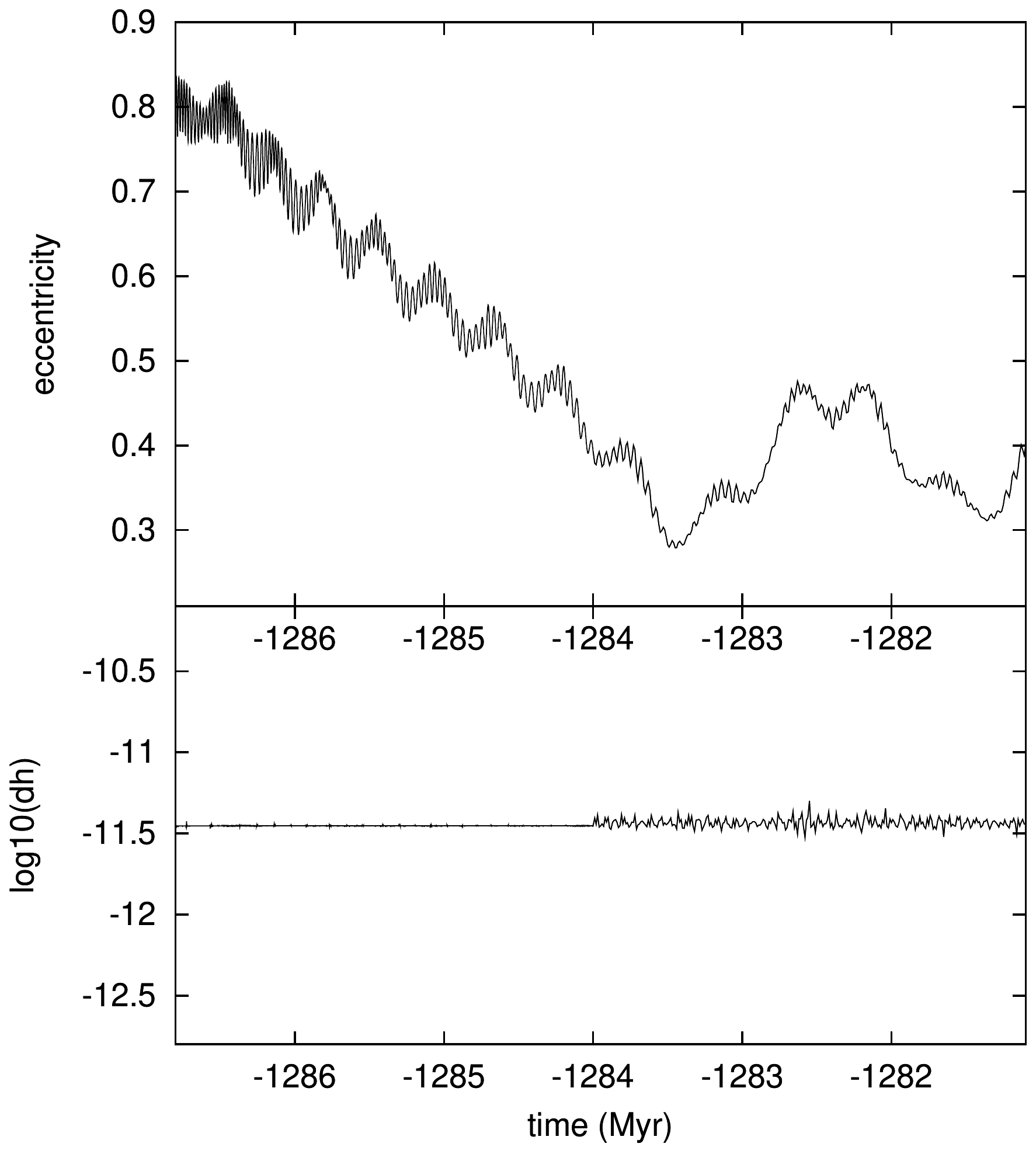} 
  \caption{Example of very high value reached by the exentricity of Mercury
  in the 
  ${\cal S}_5$ solution. The eccentricity is given versus time in Myrs (top).
  In the bottom frame, the relative variation of total energy (dh $=\abs{(h(t)-h(0))/h(0)}$) of the system is given.
  The step size of the integration is $1 \times 10^{-2}$ yr on the interval 
$[-1.284 , -1.215]$ Gyr and   $1 \times 10^{-3}$ yr over the interval 
$[-1.2868 , -1.284]$ Gyr.
  }
  \llabel{FigNn}
\end{center}
\end{figure}
}

\newcommand\figNo{
\begin{figure}[h] 
\begin{center}
 \includegraphics[width=8.5cm]{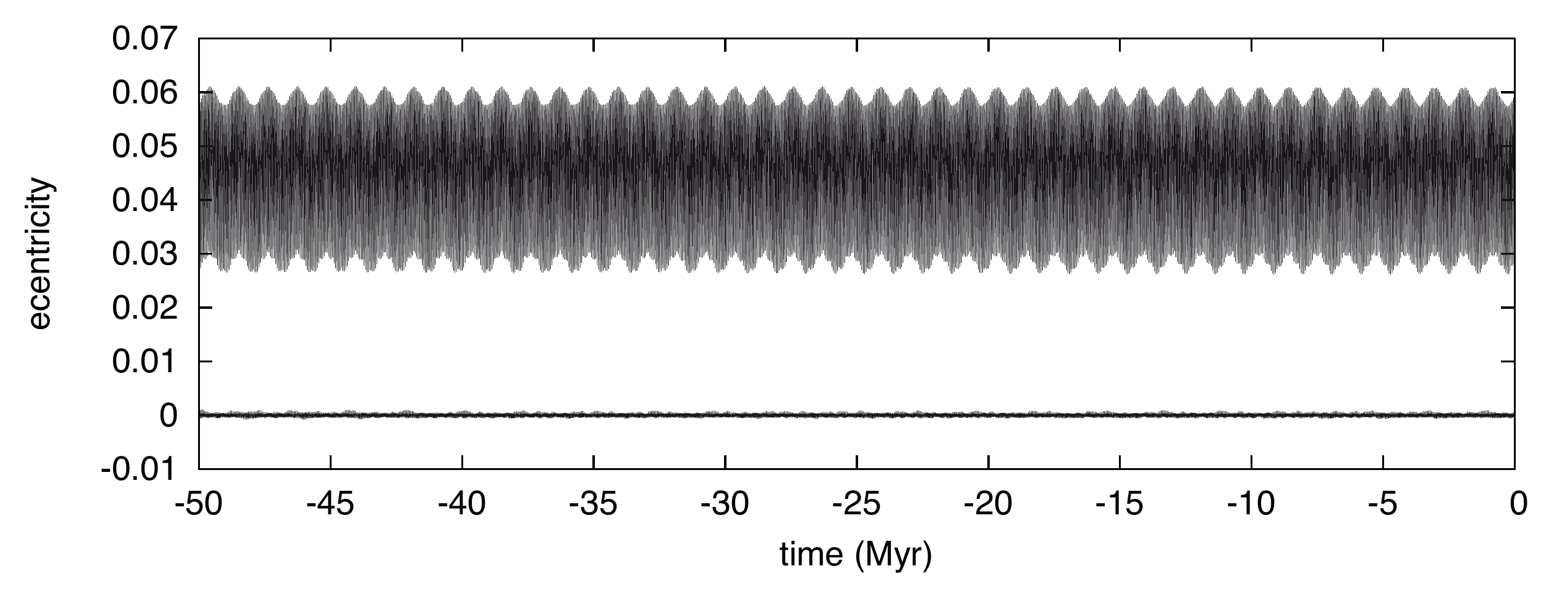} 
  \caption{Solution for the eccentricity of Jupiter (top), and residuals 
  after substraction of a  quasiperiodic approximation with 5 periodic terms (Eq. \ref{quasi5}).
  }
  \llabel{FigNo}
\end{center}
\end{figure}
}

\newcommand\figNp{
\begin{figure}[h] 
\begin{center}
 \includegraphics[width=7.5cm]{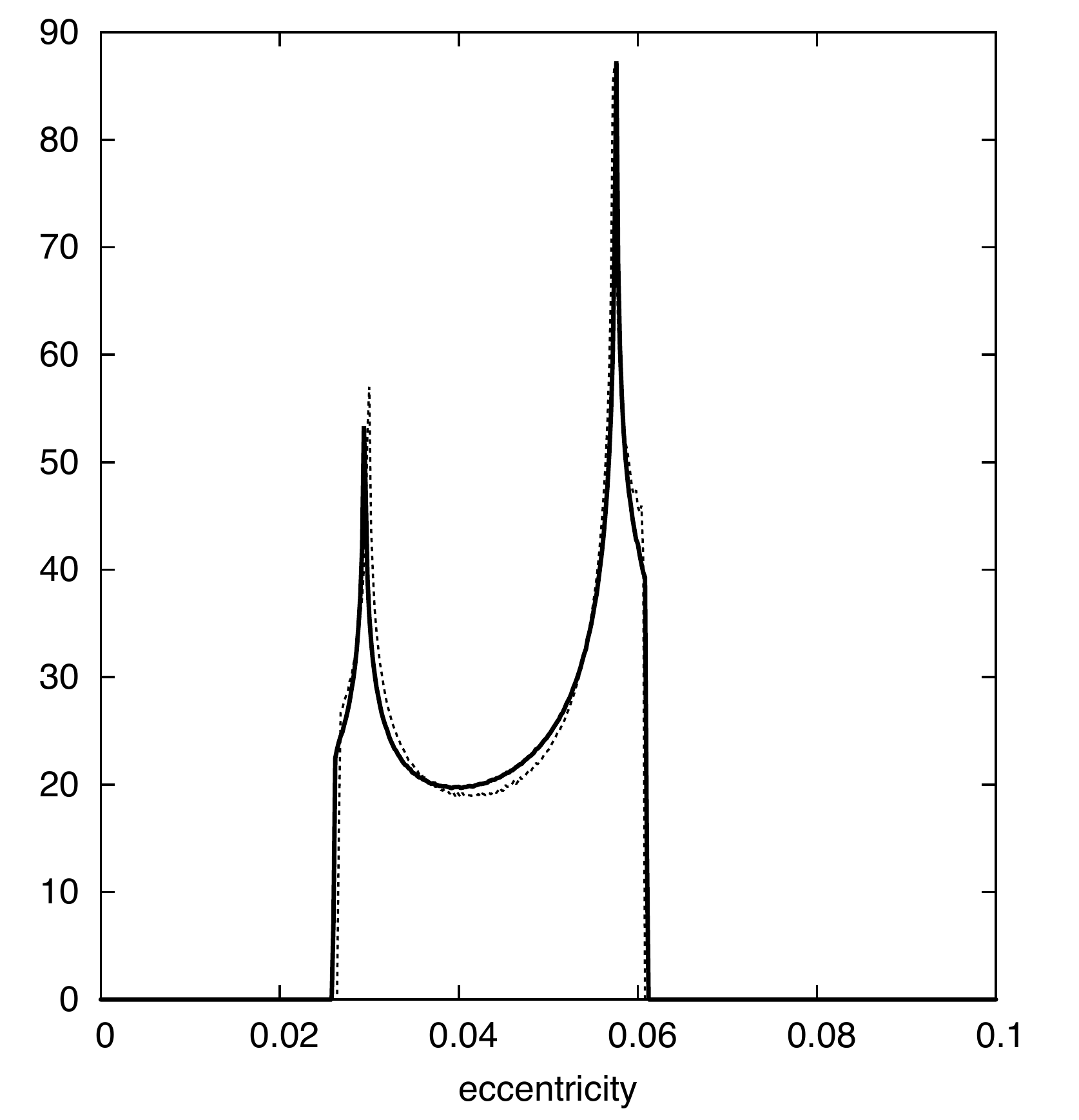} 
  \caption{Density function for the eccentricity of Jupiter (solid line) and in dotted line, the density 
  function of a quasiperiodic approximation with 5 periodic terms (Eq. \ref{quasi5}).
  }
  \llabel{FigNp}
\end{center}
\end{figure}
}

\newcommand\figNq{
\begin{figure}[h] 
\begin{center}
 \includegraphics*[width=8.5cm]{figures_pdf/figecc.pdf} 
  \caption{Secular resonance $g_1=g_5$. 
  For   initial  eccentricity of Mercury ($e0$) from $0$ to $0.95$ with step size $0.001$, 
  a numerical integration 
  of the secular system is performed over 40 Myr. The value of $g_1$ is obtained by frequency analysis  
  over 20 Myr intervals and is plotted versus initial eccentricity $e0$ in presence (a1) 
  or absence (a2) of GR.
  The horizontal line corresponds to  the secular frequency $g_5 \approx 4.257''/$yr. 
  In (b1) and (b2) are reported the maximum values of the eccentricity of Mercury reached 
  over the 40 Myr interval, with (b1) and without (b2)
  relativity. 
  With GR, $g_1$ is far from the secular resonance $g_1=g_5$ and the 
  variation of eccentricity 
  is moderate. On the opposite, when GR is not considered, the secular frequency 
  $g_1$ is smaller, 
  and as the eccentricity of Mercury increases, the orbit gets trapped into the secular resonance 
  $g_1=g_5$ that may drive 
  the eccentricity to very high values (b2).
  }
  \llabel{FigNq}
\end{center}
\end{figure}
}

\newcommand\figNr{
\begin{figure}[h] 
\begin{center}
 \includegraphics[width=8.5cm]{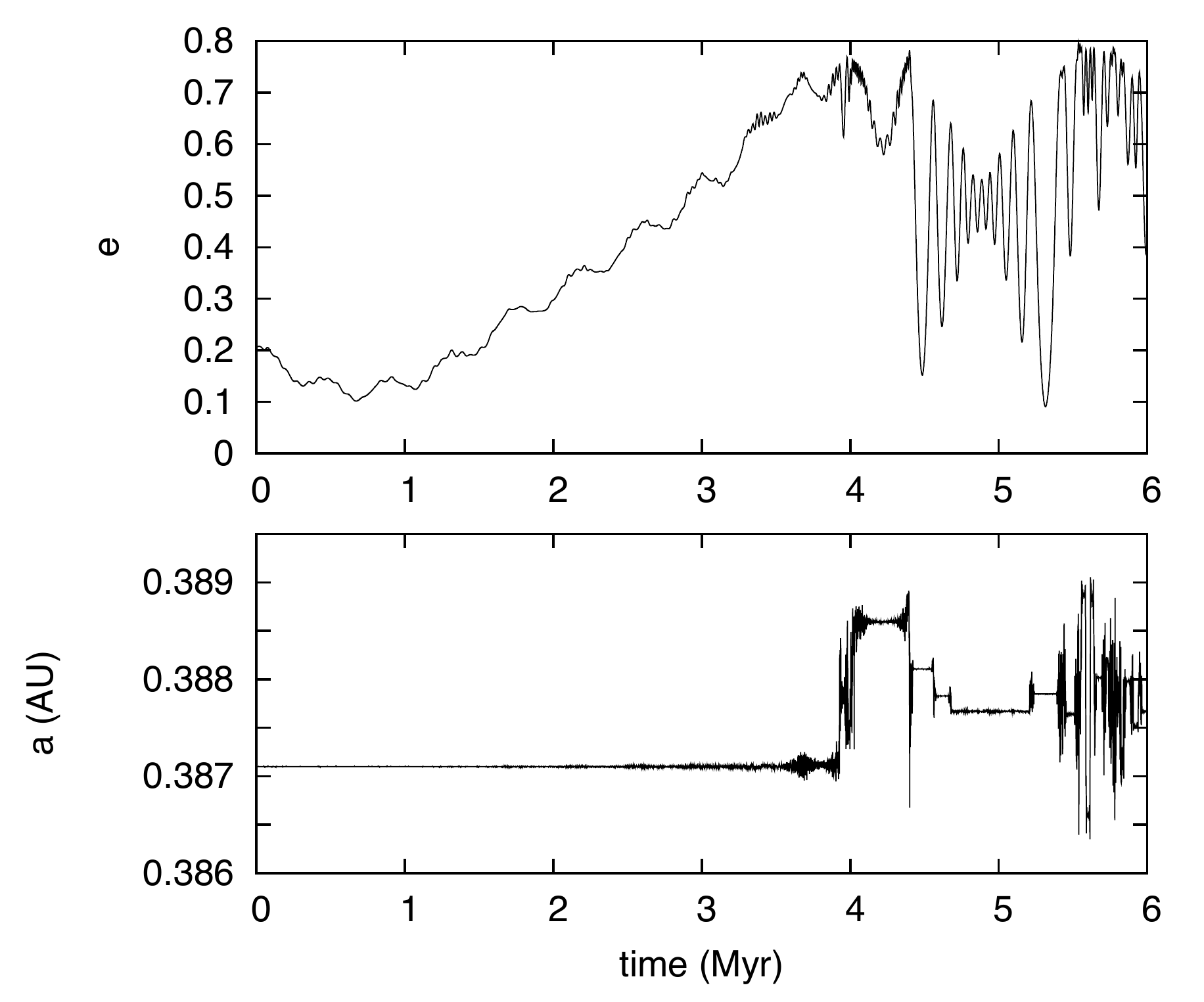} 
  \caption{Evolution of the eccentricity (top) and semi-major axis (bottom) of Mercury in a numerical integration 
  of the full Solar System with the non averaged equations with  general relativity where the PPN parameters
  $2\gamma-\beta$ has been artificially set to $-2$ instead of $1$ in the standard setting. With this setting, and 
  the present initial conditions and parameters from INPOP06 (Fienga \etal, 2008), the system is in the $g_1=g_5$ 
  secular resonance. The eccentricity of Mercury then increases to very high values in only 4 Myr, where it 
  reaches the region of mean motion resonances overlap where   chaotic variations 
  of the semi-major axis occur.
  }
  \llabel{FigNr}
\end{center}
\end{figure}
}

\newcommand\figNs{
\begin{figure}[h] 
\begin{center}
 \includegraphics[width=8.5cm]{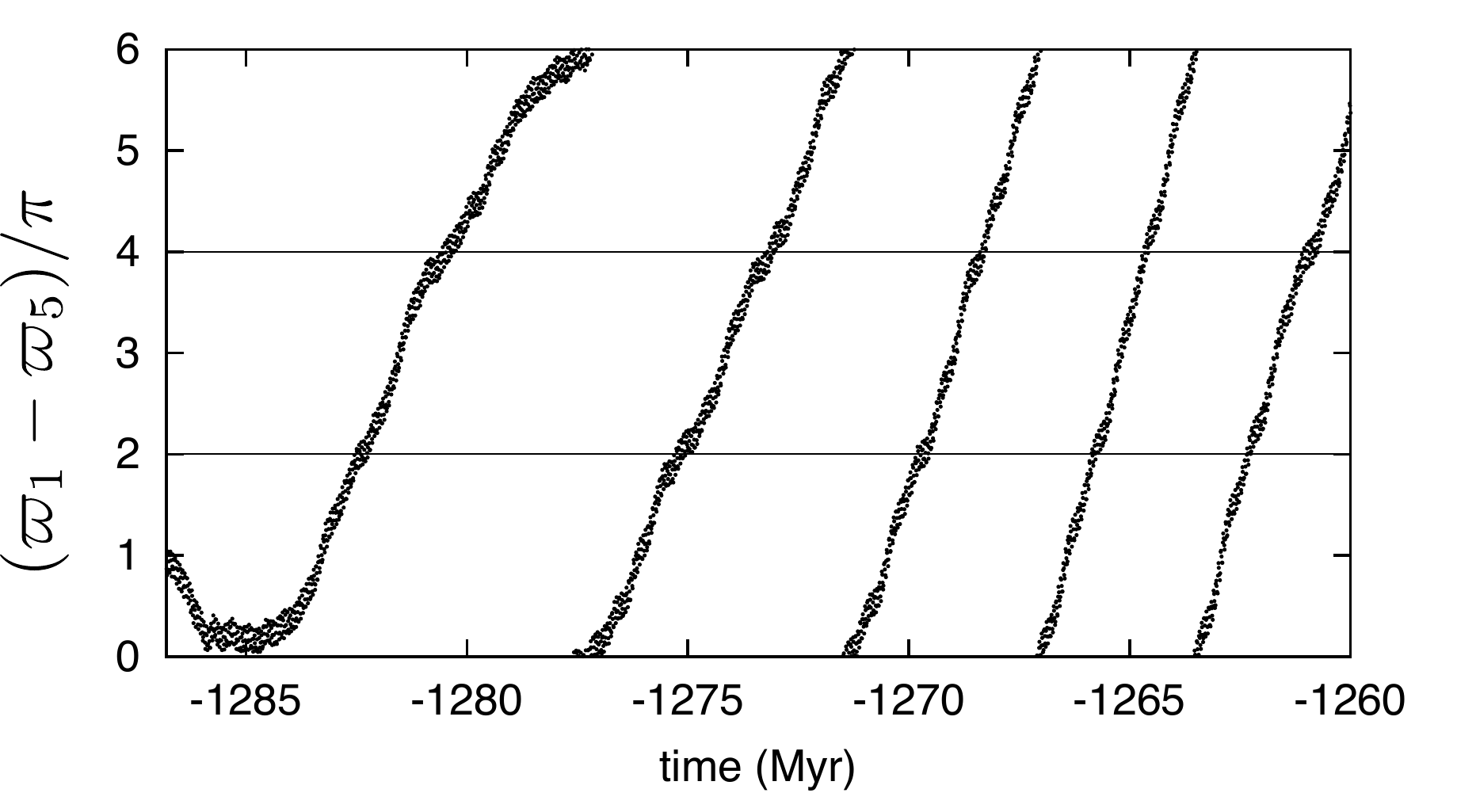} 
  \caption{Evolution of the difference of longitude of perihelion ($\varpi_1-\varpi_5$)
  of Mercury and Jupiter for the $\cS_5$ solution. The angle is plotted modulo $6\pi$ 
  in order to better visualize the trapping into resonance that occurs from $-1284$ Myr 
  to $-1286$ Myr, and corresponds to the strong increase of the eccentricity of Mercury 
  from $0.4$ to $0.8$
  (Fig. \ref{FigNn}).
  }
  \llabel{FigNs}
\end{center}
\end{figure}
}

\newcommand\tabNa{
\begin{table}[h]
\begin{tabular}{cccc}
\hline
$N$ &  $m$ &  $b_0$ & $b_1$ \\
\hline
       1  & 0.1875                    &  2.070e-03&  1.043e-03  \\
       2  & $0.02235 +0.00014\, T$   &  4.197e-04&  5.110e-06 \\
       3  & $0.01951 +0.00013\, T$  &  3.181e-04&  3.323e-06 \\
       4  & $0.06437  -0.00188\, T$  &  1.002e-03&  9.127e-05 \\
\hline 
       1  & 4.9896                    &   9.040e+00&  1.272e+00   \\
       2  & 1.5864                   &   1.492e+00&  2.377e-02 \\
       3  & 1.5803                   &   1.063e+00&  2.308e-02 \\
       4  & $4.8289   -0.1703\, T$  &   3.623e+00&  2.398e-01 \\
\hline
  \end{tabular}
  \caption{Evolution of the parameters for the Rice PDF of the inner planets Mercury, Venus, the Earth, 
  and Mars over 5 Gyr. $N$ is the index of the planet. The parameters $m$ of the Rice PDF (Eqs. \ref{eq.rice})
is given in column 2, while the coefficients $b_0, b_1$ allow one to compute the 
$\sigma$ parameter of (\ref{eq.rice})  as $ \sigma^2 = b_0 + b_1 T$, where $T$ is in Gyrs.} 
  \llabel{TabNa}
\end{table}
}

\newcommand\tabNb{
\begin{table}[h]
\begin{tabular}{ccccccccc}
\hline
 $e_{m0}$      &500   &        1000    &     1500&     2000&    3000     &    4000    &     5000 \\
\hline
  0.35 &     14&     44&     86&    127&     228&    328&    426 \\ 
  0.40 &      2&      8&     17&     37&      81&    153&    219 \\ 
  0.50 &      0&      0&      0&      1&       8&     28&     48 \\ 
  0.60 &      0&      0&      0&      0&       2&     10&     21 \\ 
  0.70 &      0&      0&      0&      0&       1&      8&     14 \\ 
  0.80 &      0&      0&      0&      0&       1&      8&     12 \\ 
  0.90 &      0&      0&      0&      0&       0&      6&      9 \\ 
\hline
  \end{tabular}
  \caption{In negative time, number of integrated solutions (out of a total of 1001 cases,  rescaled to 1000) 
  for which the maximum value reached 
  by the eccentricity of Mercury ($e_{max}$) over a given time (500Myr, 1000 Myr, 1500 Myr, 
  2000 Myr, 3000 Myr, 4000 Myr, or 5000 Myr) is above a specified value 
  ($e_{m0}$).
  } 
  \llabel{TabNb}
\end{table}
}

\newcommand\tabNc{
\begin{table}[h]
\begin{tabular}{ccccccccc}
\hline
 $e_{m0}$      &500   &        1000    &     1500&     2000&    3000     &    4000    &     5000 \\
\hline
  0.35 &    130&    341&    478&    558&       692&    763&    812 \\ 
  0.40 &     75&    249&    373&    449&       589&    684&    747 \\ 
  0.50 &     24&    118&    226&    306&       442&    552&    640 \\ 
  0.60 &     16&     76&    169&    238&       364&    476&    564 \\ 
  0.70 &     14&     67&    150&    218&       343&    454&    541 \\ 
  0.80 &     12&     63&    141&    209&       331&    442&    531 \\ 
  0.90 &     12&     61&    138&    202&       325&    441&    530 \\ 
\hline
  \end{tabular}
  \caption{When General Relativity is not taken into account, number of integrated solutions 
  in negative time
  (out of a total of 988 cases, rescaled to 1000) 
  for which the maximum value reached 
  by the eccentricity of Mercury ($e_{max}$) over a given time (500Myr, 1000 Myr, 1500 Myr, 
  2000 Myr, 3000 Myr, 4000 Myr, or 5000 Myr) is above a specified value 
  ($e_{m0}$).
  } 
  \llabel{TabNc}
\end{table}
}

\newcommand\tabNd{
\begin{table}[h]
\begin{tabular}{ccccccccc}
\hline
 $e_{m0}$      &500   &        1000    &     1500&     2000&    3000     &    4000    &     5000 \\
\hline
  0.35 &     25&     75&    128&    165&      280&    366&    427 \\ 
  0.40 &      4&     21&     38&     52&      113&    180&    243 \\ 
  0.50 &      0&      0&      0&      0&        6&     19&     33 \\ 
  0.60 &      0&      0&      0&      0&        0&      6&     10 \\ 
  0.70 &      0&      0&      0&      0&        0&      6&     10 \\ 
  0.80 &      0&      0&      0&      0&        0&      2&      8 \\ 
  0.90 &      0&      0&      0&      0&        0&      0&      2 \\ 
\hline
  \end{tabular}
  \caption{For positive time , number of integrated solutions 
  (out of a total of 478 cases,  rescaled to 1000) 
  for which the maximum value reached 
  by the eccentricity of Mercury ($e_{max}$) over a given time (500Myr, 1000 Myr, 1500 Myr, 
  2000 Myr, 3000 Myr, 4000 Myr, or 5000 Myr) is above a specified value 
  ($e_{m0}$).
  } 
  \llabel{TabNd}
\end{table}
}

\newcommand\tabNe{
\begin{table}[h]
\begin{tabular}{cccccc}
\hline
 $e_{m0}$      &500   &        1000    &     1500&     2000& 3000  \\
\hline
  0.35 &      1&     2&       4&    4&    6\\ 
  0.40 &      0&     1&       3&    3&    5\\ 
  0.50 &      0&      0&      1&    1&    2\\ 
  0.60 &      0&      0&      1&    1&    2\\ 
  0.70 &      0&      0&      1&    1&    2\\ 
  0.80 &      0&      0&      1&    1&    2\\ 
\hline
  \end{tabular}
  \caption{For the direct integration of  Newtonian equations, 
  , number of integrated solutions in negative time,
  (out of a total of 10 cases) 
  for which the maximum value reached 
  by the eccentricity of Mercury ($e_{max}$) over a given time (500Myr, 1000 Myr, 1500 Myr, 
  2000 Myr, 3000 Myr) is above a specified value 
  ($e_{m0}$).
  } 
  \llabel{TabNe}
\end{table}
}

\newcommand\tabNf{
\begin{table}[h]
\begin{tabular}{ccrrr}
\hline
       $i$ &               & $\nu_i$ ("/yr)   &   $a_i$    &   $\phi_i$ (deg)  \\
\hline
0 &                      &               &    0.045570 &            \\
1 & $g_6 - g_5  $         &   23.987561   &    0.015372 &     -82.213 \\
2 & $2g_6- 2g_5 $         &   47.975121   &    0.001849 &      15.573 \\
3 & $g_5 - g_7  $         &    1.169512   &    0.001680 &     -90.509 \\
4 & $g_6 - g_7  $         &   25.157074   &    0.000503 &    -172.658 \\
5 & $g_6 + g_7 - 2 g_5 $  &   22.818063   &    0.000491 &    -171.688 \\
\hline
  \end{tabular}
  \caption{Quasiperiodic decomposition of the eccentricity of Jupiter over 50 Myr. 
 we have  $e = a_0 + \sum  a_i  \cos (\nu_i t + \phi_i)$. The decomposition is obtained through 
 frequency analysis (Laskar, 1988, 2005). The residuals are given in figure \ref{FigNo}.
  } 
  \llabel{TabNf}
\end{table}
}

\newcommand\tabNg{
\begin{table}[h]
\begin{tabular}{ccc}
\hline
variable  & planet   &    offset \\
\hline
$k$   &  Mars    & $-100\eps$ to $+ 100\eps$ \\
$k$   &  Earth   & $-100\eps$ to $+ 100\eps$ \\
$h$   &  Venus   & $-100\eps$ to $+ 100\eps$ \\
$k$   &  Jupiter & $-100\eps$ to $+ 100\eps$ \\
$h$   &  Earth   & $-100\eps$ to $+ 100\eps$ \\
\hline
  \end{tabular}
  \caption{Offsets of the initial conditions for the 1001 integrations of 
  the secular system for the variables $k=e\cos \varpi$ and $h=e \sin \varpi$.
  The different integrations correspond to   offsets of $N\eps$ for  $N=-100,\dots, +100$
  and $\eps = 10^{-10}$,
  in a single variable, for a single planet, while the other variables are kept 
  to their nominal values.
  } 
  \llabel{TabNg}
\end{table}
}

\title{Chaotic diffusion in the Solar System. }
\author{J. Laskar\\[0.5cm]
\it
\small
 Astronomie et Syst\`emes Dynamiques, IMCCE-CNRS UMR8028, Observatoire de Paris, \\
\it\small 77 Av. Denfert-Rochereau, 75014 Paris, France 
 }

\date{{{\small \it Received December, 10; revised, February 3, 2008; accepted, February  19, 2008}}}
\begin{document}
\maketitle
\icar{\vspace{10cm}}
\icar{PAGES: 53\ ; \hspace{2cm} TABLES: 7 \ ; \hspace{2cm} FIGURES: 21}
\icar{\clearpage}
\icar{
{\large Running Title : Chaotic diffusion in the Solar System.} \\[2cm]
Corresponding author : Jacques Laskar \\
Astronomie et Syst\`emes Dynamiques, IMCCE-CNRS UMR8028, 
77 Av. Denfert-Rochereau, 75014 Paris, France\\
email : laskar@imcce.fr
\clearpage
}
\begin{abstract}
The discovery of the chaotic behavior of the planetary orbits in the Solar 
System (Laskar, 1989, 1990) was obtained using numerical integration of 
averaged equations. In  (Laskar, 1994),  these same equations are integrated 
over several Gyr and show the evidence of very large  possible increase of the 
eccentricity of  Mercury through chaotic diffusion. On the other hand, 
in the  direct  numerical integration of (Ito and Tanikawa, 2002) performed without 
general relativity over $\pm 4$ Gyr, the eccentricity of Mercury 
presented some  chaotic diffusion, but with a maximal excursion smaller than 
about 0.35. 
In the present work, a  statistical analysis is performed over 
more than 1001 different integrations of the secular equations over 5 Gyr. 
This allows to obtain for each planet, the probability  for the eccentricity 
to reach large values. In particular, we obtain that the probability of 
the eccentricity of Mercury to increase beyond $0.6$ in 5 Gyr is about 1 to 2 \%, 
which is relatively large. In order to compare with (Ito and Tanikawa, 2002), 
we have performed the same analysis without general relativity, and 
obtained even more orbits of large eccentricity for Mercury. 
In order to clarify these differences, we have performed as well a direct integration of 
the planetary orbits, without averaging, for a dynamical model that do not include   
the Moon or  general relativity with 10 very close initial conditions over 3 Gyr. 
The statistics obtained with this reduced set are comparable to the statistics of the secular 
equations, and in particular we obtain two trajectories for which the eccentricity of 
Mercury increases beyond $0.8$ in less than 1.3 Gyr and 2.8 Gyr respectively. 
These strong instabilities in the orbital motion of Mecury results from secular resonance
beween the perihelion of Jupiter and Mercury that are facilitated by the absence of 
general relativity.
The statistical analysis of the 
1001 orbits of the secular equations also provides  probability density functions (PDF) 
for the eccentricity and inclination of the 
terrestrial planets (Mercury, Venus, the Earth and Mars) 
that are very well approximated by Rice PDF. This provides a very simple representation 
of the planetary PDF over 5 Gyr. On this time-scale the evolution of the 
PDF of the terrestrial planets is found to be similar to the one of a diffusive process. 
As shown in (Laskar, 1994), the outer planets orbital elements  do not present 
significant diffusion, and 
the PDFs of their eccentricities and inclinations  are
well represented by the PDF of quasiperiodic motions with a few periodic terms.

\end{abstract}

\prep{\footnotetext[1]{E-mail address: laskar@imcce.fr}}
\icar{\clearpage}

%\prep{\tableofcontents}
%\prep{\listoffigures}
\section{Introduction}
\llabel{sec.1}

The discovery of the chaotic behavior of the planetary orbits in the Solar 
System (Laskar, 1989, 1990) was obtained using numerical integration of 
averaged equations. Since then, this chaotic behavior  has been confirmed through 
direct numerical simulation, without averaging (Quinn \etal, 1991,  Sussman and Wisdom, 
1992). More recently, integrations of more accurate planetary equations 
have been performed over  100 to 250 Myr (Varadi \etal, 2003, Laskar \etal, 2004a, b).
Due to the chaotic evolution of the system, and to the uncertainty on the
model and initial conditions, the interval of precise validity of these solutions 
is limited to about 40 Myr (Laskar \etal, 2004a), even if
some components of the solutions can be used over longer time for 
paleoclimate studies, as the 
405 kyr oscillation of the Earth eccentricity (Laskar \etal, 2004a).

Over periods of time longer than 100 Myr, it becomes hopeless to search for a precise 
solution for the orbital parameters of the inner planets (Mercury, Venus, Earth and Mars).
On the other hand, it is important to understand the possible behavior of these
solutions, and  in particular of the  possible variations of the 
action variables of the orbits (semi major axis, eccentricity and inclination).

A first study of the chaotic diffusion of the planetary orbits was made in (Laskar, 1994) 
is order to search for the maximum possible variations of the eccentricities and inclinations of 
the planets, using the secular equations.

In the present work, a statistical view of the chaotic evolution of the planetary orbits 
is described for all planets of the Solar System (Pluto is excluded). 
This study is based on the numerical integrations of 1001  different solutions 
of the averaged equations of the Solar System using very close initial conditions, 
compatible with our present knowledge. Some results concerning the orbital evolution 
of Mars were already presented in (Laskar \etal, 2004b) and we will denote this paper  
PI in the following, since in many cases, we will refer to this previous paper to avoid
duplication.

\section{The secular equations}
\llabel{sec.5.1}
\prep{\figNc}

In order to investigate the diffusion of the orbits over 5 Gyr, we will use the secular equations
of  Laskar (1990), with some small modifications.
The secular equations are obtained by averaging the equations of motion 
over the fast  angles that are the mean longitudes  of the planets.
The averaging of the equation of motion is obtained by expanding the perturbations of 
the Keplerian orbits in Fourier series of the angles, where the  coefficients 
themselves are expanded in series of the eccentricities and  inclinations.
This averaging process was conducted in a very
extensive way,  up to  second order with
respect to the masses, and through degree 5 in eccentricity and
inclination, leading to    truncated 
secular equations of the Solar System of the form

\be
{{d w}\over{dt}} = \sqrt{-1}\left\{ \Gamma \,w +
\Phi_3(w,\bar w) + \Phi_5(w,\bar w)\right\}
\label{eq.sec}
\ee
where $ w=(z_1,\dots,z_8,\zeta_1,\dots,\zeta_8)$, 
with $z_k=e_k \exp(\varpi_k)$, $\zeta_k=\sin(i_k/2)\exp(\Omega_k)$ ($\varpi_k$ is the 
longitude of the perihelion).
The $16\times 16$ matrix $\Gamma$ is
the linear  Lagrange-Laplace system, while
$\Phi_3(w,\bar w)$ and  $\Phi_5(w,\bar w)$ gather the
terms of degree 3 and 5.  

The system of equations thus obtained contains some
150000 terms, but  can be considered as a simplified system, as
its main frequencies are now the precession frequencies of the orbits
of the planets, and no longer comprise  their orbital periods. The full
system can thus be numerically integrated with a very large
step-size of 200 to 500 years. Contributions due to the
secular perturbation of the  Moon and  
general relativity are also included (see Laskar  1990, 1996, and laskar \etal, 2004b (PI)
for more details and references).

 This secular system is then simplified  and reduced to about 50000 terms, after 
 neglecting terms of very small value (Laskar  1994). Finally, a small correction 
 of the  terms of the matrix $\Gamma$ of (\ref{eq.sec}), after 
 diagonalization, is performed in order to  adjust the linear frequencies,
 in a similar way as it was  done in (Laskar  1990).
 Indeed, in the  outer planetary system, terms of higher order 
 are of some importance, but their main effect will  be  to slightly modify 
 the values of the main frequencies of the system.
 The correction that is done here is a simple way to correct for  this 
 effect.   
 With the present small adjustment, the secular solutions are  very close 
 to the  direct numerical integration  La2004 (Laskar \etal, 2004a, b)
  over about 35 Myr (PI, Figs. 15 and 16).
 As noted in PI,  this time is  about the time over which 
 the direct numerical solution itself is valid (Laskar \etal, 2004a, Figs. 20, 21), because 
 of the imperfections of the model.
 Moreover, as the step-size used in the secular equations is 200 years instead of $1.82625$ days for 
 the direct integration, over 
 very long times  the numerical noise will be smaller.
 It is thus legitimate to investigate the diffusion of the orbital motion over long times using 
 the secular equations. The major advantage, besides reducing the roundoff errors, resides 
 in the computation speed : the integration of the secular equations is 2000 times faster than 
 the integration of the non-averaged equations, and we can compute a 5 Gyr solution for the 
 Solar System in 12 hours on a Compaq alpha workstation (833 Mhz). We are thus  able to make statistics 
 over many solutions with close  initial conditions. In these computations,
 our main   limitation will be 
 the huge amount of data generated by these numerical integrations.
 
\prep{\tabNg}
\prep{\tabNb}
\prep{\tabNd}

\section{Maximum excursion}

For the present analysis, we have integrated 1001 orbital trajectories of the Solar System 
over 5 Gyr in negative time with very small variations of the initial conditions with respect 
to the nominal solution. 
The initial conditions of the secular system nominal solution  are the same 
as Table 1 and 2 from (Laskar, 1986), and are derived from the initial conditions of 
the VSOP82 solution of (Bretagnon, 1982). The phase space of the secular system (\ref{eq.sec}) is of real dimension 32. 
The various initial conditions for the 1001
cases are obtained with 
a small variation of the  initial value 
of a single secular 
variable $k=e\cos (\varpi)$ or $h=e\sin (\varpi)$, 
for a single planet,
according to Table \ref{TabNg}, leaving the other 
31 initial variables equal to their nominal values.

As we have performed 1001 numerical simulations of the whole Solar System over 5 Gyr, 
it is  impossible to display the detailed results of these integrations, and 
we have chosen here to describe the most significant features of these integrations. 
One important point, that was addressed in (Laskar, 1994) is the maximum value reached by the eccentricity 
or inclination, as a result of the chaotic diffusion of the trajectories.
In (Laskar, 1994), only 5 solutions were followed, after some small change of initial 
conditions  in every interval of 500 Myr. 
This was an economical way to obtain the maximal possible value for each parameter, 
but without any estimate of  the probability to reach these values. 
In particular, I could demonstrate  that it was possible for Mercury to reach very 
high values for its eccentricity, allowing eventually a close encounter with Venus,
while similar crossings were not reached for the other planets. 

Here we perform more conventional statistics, 
and all 1001 integrations have been followed over the whole 5 Gyr interval. As 
it was mentioned before, (Laskar, 1990, 1994), there is practically no diffusion for the 
outer planet system that behaves nearly as a quasiperiodic and regular system. 
On the opposite, there is a significant diffusion of the eccentricities and inclinations 
of the inner planets. The statistics on the maximum values reached by 
the eccentricity and inclination over different time intervals of 
50, 100, 250, 500, 1000, 2000, 3000, 4000, and 5000 Myr are displayed in figure \ref{FigNc}
for Mercury, Venus, The Earth and Mars. 

Over 50 Myr, the effect of the diffusion is not yet noticeable, and all solutions 
reach the same maximum value. This is reflected by a vertical curve 
in figure \ref{FigNc}. But beyond this time interval, significant differences 
appear. In order to make these statistics more readable, the lower part of these 
maximum graphs, corresponding to the 5 percents of the solutions with 
largest variations of the eccentricities and inclination has been enlarged 

\subsection{Discussion}

An important aspect of this study is the estimate of the probability  for the eccentricity of Mercury 
to reach high values, allowing a possible close encounter with Venus. 
Over 1 Gyr, all solutions remained with $e_{max} < 0.5 $, and only 0.1\% of  the solutions 
went beyond $0.5$ over 2 Gyr (Table  \ref{TabNb}). On the other hand, over longer time  interval, 
the chaotic diffusion allowed Mercury's eccentricity to reach very high values. In our simulations, 
 0.9\% of the solutions reached  an eccentricity of 0.9 within 5 Gyr, and 0.6\% within 4 Gyr. 
 At this point, one should remind that a  close encounter of Mercury and Venus 
 is only possible if the eccentricity of Mercury reaches values of about 0.75 (assuming an 
 eccentricity of Venus  of 0.06). We still have in our simulations about 1 \% of the solutions 
 that would allow for a close encounter with Venus.
 
 In order to test the stability of the results displayed in Table  \ref{TabNb}, 
 I have also plotted in Table  \ref{TabNd} the results obtained with a similar 
 experiment, but performed in positive time. 
 The equations are the same, but the integration step is now positive. 
 The simulation is made over 478 different orbits\footnote{This odd number of cases is purely accidental. 
 Our dedicated parallel computer broke down, and as this positive time integrations are used only as a check, 
 we considered that 478 cases were sufficient for this purpose.}, but the results are scaled 
 to 1000 for easier comparison with Table  \ref{TabNb}.
 In a same way as for a diffusive process, the results in 
 positive time are very similar to the results obtained in negative time, 
 which is what was expected. 
 
 In doing these estimates, we need to keep in mind that the equations that are integrated here 
 are the averaged equations of motions, where the disturbing function of the mutual 
 interactions of the planets is expanded in series of eccentricities and inclinations. 
  This expansion is divergent for high eccentricities. Indeed,  already in the two-body
problem, the  expansion  of the eccentric anomaly in powers of excentricity  
  has a radius of convergence  $e_m \approx 0.6627$ (see Wintner, 1947).
  Additionnally, the inverse of the distance becomes singular  at collision, that is 
  for an  eccentricity of Mercury of about $0.75$, depending on the eccentricity of Venus. 
 Moreover, in the vicinity of collisions, numerous  mean motion  resonances 
 overlap, giving rise to mean motion chaotic behavior, with possible changes in the
 semi-major axis of the planets, while these are constant in the secular system
 (see Fig. \ref{FigNr}).
 
 Nevertheless, although the truncations of series expansions involved in the construction of 
 the secular system are made without 
 estimates of the remainders, as it is usually the case in astronomy, 
 I conjecture  here that up to an eccentricity of Mercury 
 of about $0.6$, the dynamics of the full system is well represented by the 
 dynamics of the secular system. 
 Indeed, it has been observed that in many cases, 
 the range of validity of secular systems extends much beyond theoretical estimates
 (see for example Libert and Henrard, 2006).

 Moreover, my assumption is that in general, the full non averaged system is less regular 
 than the secular system.
 I can thus assume that for eccentricities of Mercury below $0.6$ the data displayed 
 in  Table \ref{TabNb}  provide good  estimates of the actual   probabilities to reach these  eccentricity  values,
 while for eccentricities above $0.6$, the data given 
 in Table \ref{TabNb}  are only lower bounds of these probabilities.

It is certainly desirable that the same statistical studies 
should be conducted with the full equations of motions, without averaging, 
although this will require a considerable amount of CPU time that is
still difficult to obtain.

On the other hand, the data displayed in Table \ref{TabNb} provide substantially 
more information than in (Laskar, 1994) where only the possibility of reaching 
high values of Mercury's eccentricity allowing for a collision with 
Venus was  demonstrated. Here we show that 
the probability to reach these high values is relatively large (about 1 to 2\%).
It thus becomes conceivable to prepare  in the near future a 
full scale numerical  simulation of the same problem with the non averaged 
equations. 

\subsection{Comparison with direct integration}
\prep{\figNm}
\prep{\figNk}
\prep{\figNl}

It would be interesting to compare with an integration of the full
equations of motion over Gyr time scale, but the author is not aware of a single
numerical integration of the kind with a comparable dynamical model, including 
the contribution of the Moon and of general relativity. The long term integrations of 
(Varadi \etal 2003, Laskar \etal, 2004a,b) use precise models but span only a few 
100 Myr. 
On the other hand, the long term integration of Ito and Tanikawa (2002) 
is performed over a few Gyr but do not include the relativistic contribution. 

\subsubsection{The integration of Ito and Tanikawa}

The numerical integration of Ito and Tanikawa (2002)  is an integration 
of the Newtonian planetary equations, without relativistic contributions. 
It does not comprise either the effect of the Moon as a separate body.
In fact, the mass of the Moon has been added to the mass of the Earth for 
five integrations ($N_{+1},N_{+2},N_{+3},N_{-1},N_{-2}$) that span 
from 3.9 to 5 Gyr in the past ($N_{-1},N_{-2}$) or in the future
($N_{+1},N_{+2},N_{+3}$). The initial conditions are taken from 
the JPL ephemeris DE245 (Standish, 1990). The integrator is 
a second order symplectic integrator (Wisdom and Holman 1991) with a step size 
of 8 days. From the figure 1 of (Ito and Tanikawa 2002) we can estimate that
the maximum relative error in angular momentum is  about  $5\times 10^{-11}$ 
for all solutions except for one for which it is about  $20\times 10^{-11}$, 
the difference being probably due to different hardware.
The maximum relative error in energy is about $6\times 10^{-9}$. 

The authors do not provide very precise details on the maximum values 
reached by the eccentricities, but mention that 
the general behavior of the eccentricity of Mercury is similar 
with the results of (Laskar, 1994, 1996), although they obtained 
a maximum eccentricity for 
Mercury of about 0.35 over $\pm 4$ Gyr. 

If we consider the numerical experiment that we conducted here 
with the secular system in negative time (Table  \ref{TabNb}) over 1001 solutions, 
we have  only 32.8 \% of the solutions that lead to an 
eccentricity of Mercury larger than 0.35,  while 71.4 \% had a maximal eccentricity larger than 
0.32, and only 26.4 \% an eccentricity larger than 0.36. There is indeed a very rapid decrease 
of the probability to reach a given eccentricity in the vicinity of 0.35. 

Due to the lack of precision 
on the maximal value reached by the eccentricity of Mercury in (Ito and Tanikawa 2002), 
we can consider that the event ($e_{max} < \sim 0.35$) reached in their simulations
has,  according to our experiment on the secular system,
a probability of about 75\% to occur. As they made 5 simulations, the 
resulting probability would be $(0.75)^5$, that is about 24\%. This is not very large, 
but not unrealistic.

Nevertheless, there is a difference in the two experiments, as our secular 
equations comprise the relativistic contributions. One should thus 
 also test the  behavior of the secular system in absence of the relativistic 
contribution.

\subsubsection{The secular system without relativity}
I have thus repeated the  previous simulations in absence 
of relativity, expecting to find a more stable system. 
But the result was the opposite, and most of the 
solutions lead to   high values of the eccentricity in 5 Gyr
(Table  \ref{TabNc}). One can see that in this case, the probability of the 
event ($e_{max} < \sim 0.35$) becomes less than 25 \%,  and the probability 
of having 5 solutions with this behavior (if they are not related)
becomes totally unrealistic with a value smaller than $(0.25)^5$, that is about $0.1 \%$. 

\prep{\tabNc}

\subsection{A new direct integration without relativity}

\prep{\figNn}

In order to clarify the situation, it was thus necessary to 
make a new direct numerical integration. 
Indeed, the probability of reaching high values of the eccentricity 
of Table \ref{TabNc} are so high that one should be able 
to obtain solutions with high eccentricity with a moderate number 
or trials. I thus decided to integrate 10 orbits, $\cS_1,\dots,\cS_{10}$, with very close initial conditions
over 3 Gyr.
\prep{\tabNe}

The model comprises the 8 major planets and the dwarf planet  Pluto with Newtonian interactions.
The Earth-Moon barycenter with the sum of the masses of the Earth and Moon  is used instead of the 
Earth. The integrator is the SABA4 symplectic integrator of (Laskar and Robutel, 2001) 
that was already used in (Laskar \etal, 2004a,b) with a step size of $2.5 \times 10^{-2}$ years. 
For a perturbed Hamiltonian of the form $H=A+\epsilon B$, using this integrator with 
a step $\tau$ is equivalent to integrate exactly a close by Hamiltonian $\tilde H$ 
where the error of method $H-\tilde H$ is of the order $O(\tau^8\epsilon) + O(\tau^2\epsilon^2)$.
The integration is conducted in extended precision on   Itanium II processors with 80 bit 
arithmetics. 

All parameters and initial conditions of the nominal solution are the same as the ones used 
in the new 
high precision planetary ephemeris INPOP06 that has been developped in our group.
The reader should refer to the associated publication (Fienga \etal, 2008) 
and  web site (www.imcce.fr/inpop) for more details.
The 10 different solutions $\cS_k$ ($k=1,\dots, 10$) have the same initial conditions 
as the nominal solution, except for a shift of $k\times 10$ cm in the initial position of the Earth.

The maximum values reached by the eccentricity of Mercury in these 10 solutions have 
been plotted in figure \ref{FigNk}. One can see that the present results differ 
substantially from the results of Ito and Tanikawa (2002), as the eccentricity can reach 
much higher values than in their simulations. In particular, for 5 solutions, the eccentricity went beyond 
$0.4$, and for two of them, $\cS_2$ and $\cS_5$, the eccentricity increased to very  high values, beyond 
$0.8$, in respectively 2776 and 1286.8 Myr (Fig. \ref{FigNn}).

In the symplectic integration, the total angular momentum is conserved. 
The relative variation of the total angular momentum is thus an estimate of the roundoff 
error in the integration. For all solutions, this error remains below $10^{-{13}}$ 
over the total length of the integration (Fig.\ref{FigNm}). 
The relative error in energy is in general below $10^{-11}$, except when the eccentricity
of Mercury reaches high values (Fig.\ref{FigNl}).  For the solution $\cS_5$ that reaches very high values 
of the eccentricity, the step size was changed to $1 \times 10^{-2}$ yr on the interval 
$[-1.284 , -1.215]$ Gyr, and then to  $1 \times 10^{-3}$ yr over the interval 
$[-1.2868, -1.284]$ Gyr  in order to achieve a  good conservation of the total energy. 
With these settings, even for an eccentricity of 0.8, the relative error of the energy remains below 
 $3 \times 10^{-{10}}$ over the whole interval of integration (Fig. \ref{FigNn}).

When we gather the results of these 10 numerical simulations in a table (Tab. \ref{TabNe}) similar to the 
tables \ref{TabNb}, \ref{TabNd}, \ref{TabNc},  one can see that  in our numerical results, 
the number of high eccentricities are 
somewhat lower, but quite comparable to the results of   our table \ref{TabNc}, especially if 
we consider the small number of  events (10) on which the present statistics are made. 

The present results, that are conducted with slightly more accurate integrations than 
the ones of Ito and Tanikawa (2002),  demonstrate clearly that the  large 
excursions of the eccentricity of Mercury described in (Laskar, 1994) 
are actually present in the full integration of the Newtonian equations. 
Moreover, the statistics obtained with the secular equations in absence of 
general relativity (Tab. \ref{TabNc}) are very close to the same statistics obtained with the direct integration
(Tab. \ref{TabNe}).

We can thus assume that the statistics obtained over 1001 integrations of the secular equations 
(Tab. \ref{TabNb}) are representative of the full system. 
Actually, as was already stated in (Laskar, 1994), we expect even that the full 
equations of motion will be slightly more unstable than the secular equations, especially for high 
eccentricities, when the overlap of  mean motion resonances will increase the chaoticity of the 
orbits. 

\subsection{The effect of relativity}
\label{sec.relat}

\prep{\figNq}
\prep{\figNs}
\prep{\figNr}

The difference of behavior of the secular system with (Table \ref{TabNb})
and without general relativity (GR) (Table \ref{TabNc}) is impressive. The results of 
the direct integrations  performed without GR 
are also in good agreement with 
the results  of the   secular equation in absence of GR. 
The contribution of GR is thus essential in order to ensure 
the relative stability of Mercury. 
It is therefore important to understand the contribution of GR in this  problem. 

The most obvious effect of GR is to increase  the perihelion frequencies of the planets 
and especially of Mercury. Indeed,  the value of the secular frequency $g_1$ related to Mercury 
is $5.59\syr$ in the vicinity of the origin with a contribution of $0.43\syr$ from general relativity.
The contribution of GR to the perihelion velocity  decreases to $0.086''$/yr for Venus, 
$0.038\syr$ for the Earth, $0.013\syr$ for Mars, and only $0.0006\syr$ for Jupiter
(e. g. Laskar, 1999, Table 4). The main frequency of the longitude of perihelion of Jupiter,
$g_5 \approx 4.257\syr$ (Laskar \etal, 2004a, Table 3) is thus not changed much by GR and the main effect 
of GR is  to increase the difference $g_1-g_5$ from $0.90\syr$ in absence of GR 
to $1.33\syr$ in presence of GR.

In order to analyse the effect of this change in the dynamics of  Mercury, I have 
performed several integrations of the secular system, with and without GR, for various values 
of the initial eccentricity of Mercury. 
The initial eccentricity of Mercury in the nominal solution is $0.2056$, while in the present experiment, 
this value is varied with a step of $0.001$ from 0 to about $0.95$ until the integration crashes rapidly. 
The integrations are computed over 40 Myr. A frequency analysis is performed in order to compute 
the secular frequencies $g_k, s_k$ of the system. For each orbit, This frequency analysis (Laskar, 1990, 1993) is 
made over slidings 20 Myr intervals with a 1 Myr offset. The values obtained for $g_1$ are reported 
in figure \ref{FigNq} with (a1) and without (a2) GR. 

The differences between the two plots are striking. With GR, when the eccentricity increases, 
$g_1$ is modified, but  mostly increases and although $g_1$  presents large variations 
resulting from the  chaotic behavior of the system, the system never gets 
trapped into the $g_1=g_5$ resonance. On the opposite,  in absence of GR, 
the $g_1$ values are smaller, and as the eccentricity increases  beyond  $e0 \approx 0.6$, 
there exists a large  chaotic zone related to the secular resonance $g_1=g_5$, 
the $g_5$ secular frequency being plotted as an horizontal line in Figs \ref{FigNq}a1,a2.
When the system is in this resonance, the eccentricity of Mercury is driven to very large excentricities, 
beyond 1 (the eccentricity is not bounded by 1 in the secular system). 
In figure \ref{FigNq}b2, one trajectory starting at about $0.5$ also increases to high values. 
Indeed, most probably, the chaotic region extends in the $0.4-0.5$ region of eccenricity, 
but it will  take some additional time  to reach the region of very strong chaos beyond $e=0.6$.

This study of the secular system is made here to explore rapidly the phase space of the system, and 
to guide our  intuition. Once we  see the importance of the $g_1-g_5$ resonance in the secular system, 
we can test its contribution on the full equations with minimal computations.

We  can now verify that the large increase of eccentricity of the solution $\cS_5$ plotted in figure 
\ref{FigNn} is actually triggered by the secular resonance $g_1=g_5$. 
In figure \ref{FigNs}, 
is plotted the difference $\varpi_1-\varpi_5$ of the longitudes of perihelion of Mercury ($\varpi_1$) and 
Jupiter ($\varpi_5$) from $-1260$ Myr to $-1286.8$ Myr. It  appears that this angle is circulating 
until the two perihelions get locked from about $-1284$ Myr to $-1286$ Myr, that is over the time interval 
that corresponds to a steady increase of Mercury's eccentricity. 
In a similar way, the solution $\cS_2$ presents as well a large increase of Mercury's eccentricity from $ 0.5$
to about $ 0.9$ in only 2.5 Myr that corresponds to a locking of the perihelions of Mercury and Jupiter 
from $-2773.5$  to $-2776$ Myr.

It should be noted that 
for Mercury and Jupiter, and contrarily to the Earth, 
the relation with the longitudes of perihelion  $\varpi_1, \varpi_5$ 
and secular frequencies $g_1, g_5$, respectively,  is straightforward.  Indeed,  in both cases, the 
secular frequency $g_k$ is clearly the leading  periodic term  of the quasiperiodic expansion 
of $z_k = e_k \exp (i \varpi_k)$ (see Table II from Laskar, 1990). 

In order to see the real effectiveness of this $g_1=g_5$ resonance, I have performed an additional 
single numerical experiment, with the full equations of motion, including GR, and starting 
with the nominal intitial conditions, that is  the initial conditions of  the planetary ephemeris 
INPOP06 (Fienga \etal, 2008), without any change of the initial eccentricity of Mercury.
But in order to set  the system into the $g_1=g_5$  resonance from the starting time of integration, I have
changed the value of the post Newtonian parameter $\beta$ to $3 $ instead of  $\beta = 1 $ in standard GR,
while keeping $\gamma=1$ (Will, 2006).
The contribution factor of perihelion shift $2\gamma -\beta$ is then $-2$ instead of $+1$. The effect 
of this modified relativity is thus to  artificially decrease the secular frequency $g_1$ 
by $0.86\syr$, instead of adding $0.43\syr$ as in standard GR. The system  is thus 
from the begining in the $g_1=g_5$ resonance, and the effect is immediate as 
Mercury's eccentricity increases beyond 0.7 in less than 4 Myr (Fig. \ref{FigNr}). 
Once in this region of high eccentricity, strong chaotic behavior due to short period 
resonances induces significant changes in the semi-major axis of Mercury (Fig. \ref{FigNr}).

\section{Density functions}
\prep{\figNa}
\prep{\figNb}

The maximal possible value of the planet eccentricity is important for the
analysis of the system stability, but 
it concerns the most exceptional orbits of the system, and not necessarily 
the most probable behavior of the Solar System over its age.
For the understanding of  the 
general behavior of the Solar System, the density function for the eccentricity and 
inclination of the planets is a complementary information that 
can be very valuable for the analysis of many physical parameters 
during the evolution of the Solar System. It  was for example useful for the 
understanding of the capture of Mercury into the 3/2 spin orbit resonance 
(Correia and Laskar, 2004), or for the analysis of the past climate evolution of
Mars (Laskar \etal, 2004b). Here we have systemized the approach elaborated
in PI for all the planets of the Solar System, with   statistics over 
1001 different orbits in order to obtain a complete  statistical view of the
variations of the orbital parameters of the Solar System.

Because of the chaotic behavior of the system, we know that it will never be possible to retrieve 
precisely the past (or future) orbital evolution of the Solar System over more than a 
few tens of millions of years (Laskar, 1989, 1990). On the other hand, beyond 
about 250 Myr one can obtain  very smooth density functions for the eccentricity 
of the planets that will tell us the main general behavior of the orbits. 
Moreover,   these density functions are  some of the  
only accurate informations one can obtain  beyond a few 100 Myr. 

As in PI, we have divided here the time interval in 250 Myr intervals, and   
statistics are done over each 250 Myr interval, using the set of 1001 orbital solutions in 
negative time with the initial conditions of Table \ref{TabNg}, for 
which the output has been recorded with a 1000 yr step size. 
The first  250 Myr interval is discarded, as the randomization due to the chaotic evolution 
of the system has not yet taken place (see PI for a more complete 
analysis of Mars eccentricity solution), and the 
normalized 
probability distributions functions (PDF), for the eccentricities and inclinations of all the planets are displayed 
in figures \ref{FigNa} and \ref{FigNb}.

\prep{\figNd}
\prep{\figNeb}
\prep{\figNe}
\prep{\figNf}

\subsection{Discussion}

The difference in the density functions for the eccentricity (or inclination ) of the inner 
(Mercury, Venus, Earth, Mars) or outer planets (Jupiter, Saturn, Uranus, Neptune)
is striking on these plots. First of all, the shape of the PDF is different. 
For the inner planets, the PDF  is similar to a Gaussian curve, although 
slightly different. In particular, all the curves have zero values for 
$e=0$, with a positive, linear slope, while they have some long tail 
for large values of the eccentricity.

On the opposite, for the outer planets the density curves are strictly confined between 
two non zero values, and 
the curves have two peaks close two the minimum and maximum value of the eccentricity 
(resp. inclination).

Another striking feature is the fact that for the inner planets, we see clearly that 
numerous curves are displayed in each plot, as the density curve 
is different for each time interval of 250 Myr. 
This is the result of a significant diffusion that occurs in the eccentricity 
and inclination (Laskar, 1994), while for the outer, as the diffusion is 
practically non existent, all the 19 density curves that are actually 
plotted in figures \ref{FigNa} and \ref{FigNb} are virtually identical and 
superpose nearly exactly.

The difference between the two kinds of PDF will even be more  striking 
in the next section, as we will attempt to associate to each PDF
the density function of a simple dynamical system. For the inner planets, we found that 
the densities were very well modelized by a Rice distribution, 
(Rice, 1945) that is characteristic of either the modulus of a random 
walk in two dimensions, or alternatively to a quasiperiodic signal with 
noise (Rice 1945). We will denotes these density functions 
Rice densities. 

On the opposite, the outer planets density function is in fact representative of a density function of 
a quasiperiodic signal with a moderate number of periodic terms. 
We will call such a density function a quasiperiodic density.

In  the next section, we will examine more closely these two different densities.

\section{Probability density functions}

\subsection{Rice densities}

The Rice distribution is a continuous probability distribution with density function
\be
f_{\sigma, m}(x) = \Frac{x}{\sigma^2} \exp \left( - \Frac{x^2+m^2}{2 \sigma^2}\right) 
I_0\left(\Frac{x m}{\sigma^2}\right)
\label{eq.rice}
\ee

where $I_0(z)$ is the modified Bessel function of the first kind obtained with its 
series expansion

\be
I_0(x) = \sum_{n=0}^\infty \left(\Frac{x}{2}\right)^{2n}\Frac{1}{n!^2}
\ee

This distribution  is obtained for example for the modulus of $z= x + i y $ where $x,y$ are two Gaussian 
independent variables with variance $\sigma^2$ and mean $a_x,a_y$, with $m =\sqrt{a_x^2+a_y^2}$.

These PDF could  thus be well suited for  the eccentricity, if the rectangular variables 
$h=e\sin(\varpi)$ and $k=e\cos(\varpi)$ 
become practically Gaussian random variables because of the chaotic diffusion. 

A Rice PDF  has been successfully fitted  to the eccentricity and inclination of the inner planets 
for each of the 250 Myr  time intervals. It is impossible to display here graphics demonstrating the quality 
of the adjustment for all of these intervals,  so we will reproduce  here only the most representative cases, 
for the eccentricity of Mercury (Fig.\ref{FigNd}), Venus  (Fig.\ref{FigNeb}), Earth  (Fig.\ref{FigNe}), and Mars  (Fig.\ref{FigNf}), 
at 3 different epochs, while all cases can be 
summarized in Table \ref{TabNa}.

It is indeed remarkable that the PDF of the eccentricity and inclination of the planets are 
such smooth functions. It is as well remarkable that they are so well approximated 
by simple 2 parameters PDF ($f_{\sigma, m}$) of a single curve  family with 
relatively simple expression (\ref{eq.rice}).

\subsection{Diffusion over  5 Gyr}
\label{dif5}
\prep{\tabNa}

 As the PDF of eccentricity and inclination are well approximated for various epochs 
 by  Rice PDF, we have derived by least square fit, the values of the parameters  
 $m,\sigma $, of these PDF for all variables and all time intervals of 250 Myr. 
 Except for Mars, the $m$ value (related to the mean) do not present large variations
 over time (Figs. \ref{FigNg}, \ref{FigNhb}, \ref{FigNh}, \ref{FigNi}). On the opposite, 
 the standard deviation parameter $\sigma$ increases  with time, 
 and we have a nearly linear relation 
 \be
 \sigma^2 = b_0 + b_1 T
 \label{eq.sig}
 \ee
 similar to a  diffusion process. In Table \ref{TabNa} are gathered 
 the fitted values of $m, \sigma^2$ over 5 Gyr.

We have thus in equation  \ref{eq.rice}  and  Table \ref{TabNa}  some very simple formulas that 
will allow one to represent the PDF of the eccentricity and inclination 
of the inner planets over very long times,
of several Gyr. Quite remarkably, although it is impossible to predict the precise evolution 
of the individual trajectories, we are able to 
give very simple expressions that fit well with the observed
eccentricity and inclination  PDF of the inner planets.

These formulas can then be used for the analysis of the past or future behavior of the Solar System.
They could be used for example to compute an estimate of the capture probability of 
Mercury in the 3/2 resonance without requiring heavy numerical computations 
(Correia and Laskar,  2004). 

\prep{\figNg}
\prep{\figNhb}
\prep{\figNh}
\prep{\figNi}

We have used here a different PDF than in (Laskar \etal, 2004b), where 
a model of a random walk with an absorbing edge at zero was used for Mars eccentricity. 
In fact, the results obtained with the two PDF are very similar in this case, 
but we prefer the Rice formulation that can be more easily interpreted. Moreover, 
we see here that the evolution of the system is well describe by a diffusive process 
for the rectangular coordinates $k, h$ that behave like random Gaussian variables 
on long time scale. Actually, the variance of the eccentricity evolves linearly 
with time. This is somewhat different from the observation made 
in (Laskar, 2004) for Mars eccentricity, but this is because here we are searching 
for models that fit more generally with the behavior of all the inner planets, for 
both eccentricity and inclination.

\section {Outer planets}
\prep{\tabNf}
\prep{\figNj}

\prep{\figNo}
\prep{\figNp}
The PDF of the outer planets (Figs. \ref{FigNa}, \ref{FigNb}) are very different from the PDF 
of the inner planets.  Moreover, as it was already stated, 
the diffusion is practically inexistent, and all PDF curves over the different time intervals 
are virtually identical. The shape of these outer planets PDF can be understood as 
the PDF of some quasiperiodic functions of time with a small number of harmonics. 
Indeed, for a simple periodic function, $g(t) = \sin(t)$, 
the PDF is 
\be
f(x) = \Frac{1_{]-1,1[}(x)}{\pi\sqrt{1-x^2}}
\label{quasi5}
\ee
where $1_{]-1,1[}(x)$ is the characteristic function\footnote{
The characteristic function $1_{]a,b[}(t)$ of the interval $]a,b[$ is defined 
as $1_{]a,b[}(t)=1$ for $t\in ]a,b[$, and $1_{]a,b[}(t)=0$ if $t\notin ]a,b[$.}
of the interval $]-1,1[$ (Fig.\ref{FigNj}).

One can now understand that when for $g(t)$, instead of a single sine term, we have several periodic terms, 
the PDF of $g(t)$ will become slightly distorted with respect to a pure sine function, and 
we will recover the specific form of the PDF of the outer planets eccentricities and inclinations 
(Figs. \ref{FigNa}, \ref{FigNb}). In order to illustrate this, we will approximate the eccentricity of Jupiter 
with a few periodic terms, obtained through frequency map analysis (Laskar, 1990, 2005).

With a frequency analysis of the eccentricity of Jupiter over 50 Myr, we obtain a quasiperiodic approximation of the 
eccentricity with 5 periodic terms  on the form 
\be
e = a_0 + \sum_{i=1}^5  a_i  \cos (\nu_i t + \phi_i)
\ee
where the values of $a_i,\nu_i, \phi_i$ are given in  table  \ref{TabNf}. The eccentricity of Jupiter over 50 Myr 
is plotted in figure \ref{FigNo}, as well as its difference with this quasiperiodic approximation. 
The density function obtained with this  quasiperiodic approximation is plotted in 
figure \ref{FigNp} in dotted line together with the density function of the full 
numerical integration (in full line). 
The two PDF are very close. It will be the same for the other  outer planets. For these planets with 
a motion that is very close to quasiperiodic, the quasiperiodic decomposition obtained over 
50 Myr as in (Laskar, 1990) provides in fact a very good representation of the orbit over 
several Gyr. It is thus not necessary to reproduce these expressions here, and we would rather 
provide  in a forthcoming publication some full and accurate quasiperiodic representations for the motion 
of the outer planets, in agreement with our latest adjustment of the planetary 
ephemeris to observations (Fienga \etal, 2007).

\section{Conclusions}
\llabel{sec.6}
In (Laskar 1994), I showed the possibility of a very large increase of the eccentricity of 
Mercury, allowing for a close encounter  or a collision with Venus. But 
in this work, there was no 
estimate of the probability for such an event to take place within a few Gyr. 
Here, such an estimate is given, by the extensive study of  more than 1001 orbits. 
Indeed, it is found that the probability for Mercury's eccentricity to exceed 0.6 
within 5 Gyr is about 1 to 2 \% which can be considered as a large value for such an important event. 

When the contributions of general relativity and the Moon are not taken into account, this 
probability increases in a large amount and in the present  numerical simulations, nearly half 
of the orbits went beyond 0.6 in less than 4 Gyr. 
As this appear to be in contradiction with the numerical results of (Ito and Tanikawa 2002), 
I have also performed a direct numerical simulation of the Newtonian equations, with 10 
nearby initial conditions, without the Moon 
or general relativity over  3 Gyr, and found results that are in good agreement with 
the results of the integration of the secular system. In particular,  two 
orbit were found for which the eccentricity of Mercury rises to very large values, beyond 0.8, 
thus allowing for a close encounter or a  collision with Venus.

This  direct numerical simulation is performed with slightly better accuracy 
than  the  numerical integration of (Ito and Tanikawa 2002). 
One should wonder about the reason of such a different behavior in the present computations  and 
in the simulation of  (Ito and Tanikawa 2002). 
The most probable reason for this is that  
the two integrations do not have the same initial conditions or model. The solutions of 
(Ito and Tanikawa 2002) may evolve in a slightly  more regular region of the phase space.
On the other hand, 
the secular equations are in very good agreement with the present direct integrations. 
This  comfort their reliability and usefulness as for the secular equations, it was possible to perform an 
extended  statistical study over 1001 solutions. 
One could  perform additional direct numerical  integrations in order to verify the probability 
law for the excursion of Mercury's eccentricity, but  this is of limited interest if 
it concerns  the model of pure Newtonian equations, and not the full model including general 
relativity. 

Indeed, as we have demonstrated here in section \ref{sec.relat},
the contribution of general relativity  changes in a considerable 
manner the behavior of the Solar System dynamics. 
Indeed, in absence of GR, the  secular frequency of the perihelion of  Mercury $g_1$ decrease by 
$0.43\syr$, and becomes closer to the secular frequency of the perihelion of Jupiter $g_5$. 
As the eccentricity varies under the secular planetary perturbations, the system can enter 
into the $g_1=g_5$ secular resonance that  can drive  Mercury's eccentricity to very high 
values, beyond $0.8$, where additional short period chaotic behavior occurs, inducing 
changes of semi-major axis of the planet (Fig. \ref{FigNr}).
This is indeed the mechanism that is present in our  two  solutions 
of the Newtonian equations $\cS_2, \cS_5$  for which Mercury's eccentricity increased 
beyond $0.8$ (Fig. \ref{FigNs}).

An important result of the present paper is to show this possibility of  very large increase 
of Mercury's eccentricity, beyond 0.8,  allowing for a possible 
collision with Venus. It was actually possible to see that 
with such a large value of the eccentricity, the planet 
entered a region of mean motion resonances and chaos inducing changes in its   
semi-major axis.

It is clearly desirable  to conduct now  a full scale numerical experiment  
with the complete equations including general relativity and the contribution of the 
Moon, in order to provide some precise results on the chaotic behavior 
and possible evolution of our Solar System.  We are indeed planning such a numerical  
study for the near future.

Apart from demonstrating the possibility of very large values for the 
eccentricity of Mercury, as was forecasted in (Laskar, 1994), I provide here 
probability density functions (PDF) for the eccentricities  and inclinations of 
the terrestrial planets. 
It is in fact quite remarkable that the PDF of all terrestrial planets, when computed over 
long time, beyond about 500 Myr, become very smooth functions that are   well 
approximated by a very simple two parameters function, namely the Rice distribution.  
This provides very compact  formulas for the evolution of these PDF over time for all 
the terrestrial planets. The evolution of the PDF with time is similar to the one of a diffusion process
with a linear increase of the variance variable $\sigma^2$ with time.

For the outer planets, we have seen here that the chaotic diffusion is extremely small,
without practical change of the PDF over time. The PDF being well approximated by 
the PDF of a quasiperiodic approximation of the solution. 
In the   full equations of motion of the outer Solar System, 
there is some chaotic behavior that has been described 
(Sussman and Wisdom, 1992, Murray and Holman, 1999, Guzzo, 2005), but 
although a more detailed analysis should be made, It seems that 
the diffusion induced by this intrinsic 
chaotic behavior resulting from mean motion interactions should be  smaller
than the diffusion induced by the chaotic behavior of the inner Solar System that is already 
included in the present work. It is thus doubtful that the PDF for the outer 
planets that have been computed here will be much changed by the consideration of 
the full dynamics of the outer Solar System.

More generally, the PDF that are given here are the PDF obtained with the secular equations. Nevertheless, 
they should be very stable with respect to small changes in the 
model or in the involved parameters, and  we can conjecture  
as well that they are very close to the PDF  of the full system.

\subsection*{Acknowledgments} 
The assistance of Mickael Gastineau is greatly acknowledged.
A large part of the computations were made at IDRIS-CNRS. This work also benefited from support 
from PNP-CNRS, and CS from Paris Observatory.

%%%%%%%%%%% teste
\setlength{\leftmargin}{-2.cm}
\setlength{\bibindent}{0.3cm}
\itemindent=-2cm
%%%%%%%%%%%% non teste
\parindent=0pt
\setlength{\parindent}{-2cm}
\itemsep=0 pt 
\parsep=0pt
%%%%%%%%%%%%%%%%%%%%%% format of biblio with openbib
\makeatletter \renewcommand\@biblabel[1]{} \makeatother   
%%%%%%%%%%%%%%%%% Biblio format Icarus %%%%%%%%%%%%%%%%%%%%%%%%%%%%%%%%%%%%%%%%%%%%%%%%
\def\bib#1#2#3#4#5#6#7{\bibitem{#1} {#2}\  {#3}.\ {#4}. {{\it #5}} {{\bf #6,}} {\ #7}.  \par }
\def\bibx#1#2#3#4#5#6#7{\bibitem{#1} {#2}\  {#3}.\ {#4} \  {{\it #5}} {{\bf #6,}} {\ #7}.  \par }
\def\bibB#1#2#3#4#5{\bibitem{#1} {#2}\  {#3}. {{\it #4}.} {(#5, #3)}. \par }
\def\bibll#1#2#3#4#5#6#7{\bibitem{#1} {#2}\  {#3}.\ {#4}. {{\it #5}}, {#7}.\par }
\def\bibb#1#2#3#4#5{\bibitem{#1} {#2}\  {#3}.\ {{\it #4,}} {#5}}.
\def\bibC#1#2#3#4#5#6{\bibitem{#1} {#2}\  {#6}.\ :{\ #3},\ {\ #4},\ {{\it #5}}\ {{\it #6}}.  \par }
%%%%%%%%%%%%%%%%%%%%%% end format of biblio 

%%%%%%%%%%%%%%%%%%%%%%%%%%% arret pour le preprint 
\prep{\end{document}}

\setcounter{figure}{0}
\setcounter{table}{0}

\clearpage
\section*{Tables and Figures}
\clearpage

\tabNg
\clearpage

\tabNb
\clearpage

\tabNd
\clearpage

\tabNc
\clearpage

\tabNe
\clearpage

\tabNa
\clearpage

\tabNf
\clearpage

\figNc
\clearpage

\figNm
\clearpage

\figNk
\clearpage

\figNl
\clearpage

\figNn
\clearpage

\figNq
\clearpage

\figNs
\clearpage

\figNr
\clearpage

\figNa
\clearpage

\figNb
\clearpage

\figNd
\clearpage

\figNeb
\clearpage

\figNe
\clearpage

\figNf
\clearpage

\figNg
\clearpage

\figNhb
\clearpage

\figNh
\clearpage

\figNi
\clearpage

\figNj
\clearpage

\figNo
\clearpage

\figNp
\clearpage

\end{document}